\documentclass[iop, twocolumn]{aastex63}
\usepackage{verbatim,subfigure}
\hypersetup{hidelinks}
\shorttitle{APOLLO-Timing}
\shortauthors{Wen et al.}
\usepackage{ulem}
\usepackage{booktabs}
\usepackage{amsmath,amssymb,bm}
\usepackage[para]{threeparttable}
\begin{document}

\title{APOLLO-Timing: Annealed Phase–Orbit Lattice Linking across Observations for Binary Pulsar Timing}

\correspondingauthor{D. Zhao}
\email{zhaode@xao.ac.cn}

\author{Y. R. Wen\href{https://orcid.org/0009-0006-4212-3801}
{\includegraphics[scale=0.04]{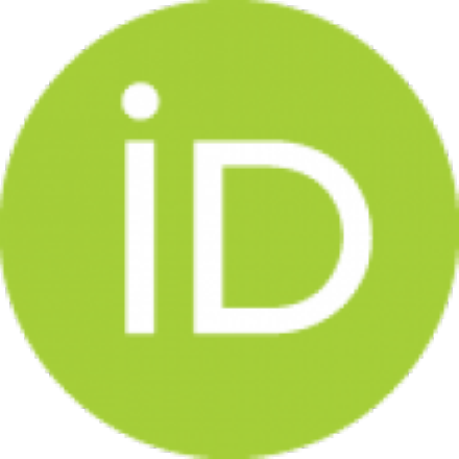}}}
\affiliation{State Key Laboratory of Radio Astronomy and Technology, National Astronomical Observatories, CAS, Xinjiang Astronomical Observatory, 150 Science 1-Street, Urumqi, Xinjiang, 830011, China}
\affiliation{University of Chinese Academy of Sciences, 1 Yanqihu East Road, Huairou District, Beijing, 101408, China}
\affiliation{Xinjiang Astronomical Observatory, Chinese Academy of Sciences, 150 Science 1-Street, Urumqi, Xinjiang 830011, China}
\affiliation{Xinjiang Key Laboratory of Radio Astrophysics, 150 Science 1-Street, Urumqi, Xinjiang, 830011, China}
\author{D. Zhao\href{https://orcid.org/0009-0007-8062-1454}{\includegraphics[scale=0.04]{ORCIDiD.pdf}}}
\affiliation{State Key Laboratory of Radio Astronomy and Technology, National Astronomical Observatories, CAS, Xinjiang Astronomical Observatory, 150 Science 1-Street, Urumqi, Xinjiang, 830011, China}
\affiliation{Xinjiang Astronomical Observatory, Chinese Academy of Sciences, 150 Science 1-Street, Urumqi, Xinjiang 830011, China}
\affiliation{Xinjiang Key Laboratory of Radio Astrophysics, 150 Science 1-Street, Urumqi, Xinjiang, 830011, China}

\author{N. Wang\href{https://orcid.org/0000-0002-9786-8548}{\includegraphics[scale=0.04]{ORCIDiD.pdf}}} 
\affiliation{State Key Laboratory of Radio Astronomy and Technology, National Astronomical Observatories, CAS, Xinjiang Astronomical Observatory, 150 Science 1-Street, Urumqi, Xinjiang, 830011, China}
\affiliation{Xinjiang Astronomical Observatory, Chinese Academy of Sciences, 150 Science 1-Street, Urumqi, Xinjiang 830011, China}
\affiliation{Xinjiang Key Laboratory of Radio Astrophysics, 150 Science 1-Street, Urumqi, Xinjiang, 830011, China}

\author{J. P. Yuan\href{https://orcid.org/0000-0002-5381-6498}{\includegraphics[scale=0.04]{ORCIDiD.pdf}}}
\affiliation{State Key Laboratory of Radio Astronomy and Technology, National Astronomical Observatories, CAS, Xinjiang Astronomical Observatory, 150 Science 1-Street, Urumqi, Xinjiang, 830011, China}
\affiliation{Xinjiang Astronomical Observatory, Chinese Academy of Sciences, 150 Science 1-Street, Urumqi, Xinjiang 830011, China}
\affiliation{Xinjiang Key Laboratory of Radio Astrophysics, 150 Science 1-Street, Urumqi, Xinjiang, 830011, China}

\author{\href{mailto:1983wangjingbo@163.com}{J. B. Wang}}
\affiliation{Lishui University, Lishui, China}

\author{J. T. Bai\href{https://orcid.org/0000-0002-1052-1120}{\includegraphics[scale=0.04]{ORCIDiD.pdf}}}
\affiliation{Institute for Gravitational Wave Astronomy, Henan Academy of Sciences, Zhengzhou 450046, Henan, People’s Republic of China}

\author{\href{mailto:chenzihao@xao.ac.cn}{Z. H. Chen}}
\affiliation{State Key Laboratory of Radio Astronomy and Technology, National Astronomical Observatories, CAS, Xinjiang Astronomical Observatory, 150 Science 1-Street, Urumqi, Xinjiang, 830011, China}
\affiliation{Xinjiang Astronomical Observatory, Chinese Academy of Sciences, 150 Science 1-Street, Urumqi, Xinjiang 830011, China}
\affiliation{Xinjiang Key Laboratory of Radio Astrophysics, 150 Science 1-Street, Urumqi, Xinjiang, 830011, China}
\affiliation{University of Chinese Academy of Sciences, 1 Yanqihu East Road, Huairou District, Beijing, 101408, China}

\author{W. M. Yan\href{https://orcid.org/0000-0002-7662-3875}{\includegraphics[scale=0.04]{ORCIDiD.pdf}}}
\affiliation{State Key Laboratory of Radio Astronomy and Technology, National Astronomical Observatories, CAS, Xinjiang Astronomical Observatory, 150 Science 1-Street, Urumqi, Xinjiang, 830011, China}
\affiliation{Xinjiang Astronomical Observatory, Chinese Academy of Sciences, 150 Science 1-Street, Urumqi, Xinjiang 830011, China}
\affiliation{Xinjiang Key Laboratory of Radio Astrophysics, 150 Science 1-Street, Urumqi, Xinjiang, 830011, China}

\author{D. Li}
\affiliation{New Cornerstone Science Laboratory, Department of Astronomy, Tsinghua University, Beijing 100084, China}
\affiliation{National Astronomical Observatories, Chinese Academy of Sciences, Beijing 100101, China}

\author{C. C. Miao}
\affiliation{College of Physics and Electronic Engineering, Qilu Normal University, Jinan 250200, China}

\author{P. Wang}
\affiliation{National Astronomical Observatories, Chinese Academy of Sciences, Beijing 100101, China}

\author{W. W. Zhu}
\affiliation{National Astronomical Observatories, Chinese Academy of Sciences, Beijing 100101, China}

\begin{abstract}
Sparse observations of binary pulsars can leave the orbit and the pulse count across observing gaps uncertain. We present APOLLO-Timing, a Bayesian phase--orbit lattice method that fits timing parameters together with integer pulse offsets between observing sessions. To validate candidate phase connections, we refit the timing model at fixed pulse numbers using the Python pulsar timing package PINT. We demonstrate the method on PSR J1912$-$1215, discovered in the CRAFTS survey. A preliminary Keplerian fit to variations in the apparent spin period and its derivative did not phase-connect 96 times of arrival (TOAs) spanning 1438.07 days in 23 sessions. Starting from its position, spin frequency $F_0$, and first time derivative $F_1$ estimated in that fit, with dispersion measure fixed, APOLLO-Timing recovers the orbit and a unique best-supported phase-connected solution without supplied orbital parameters. A repeat full-data blind search recovers the same pulse counts in 914.0 s, including the final fit. Setting the initial $F_1$ to zero in the 17-, 18-, and 19-session subsets also recovers the reference path. The solution gives $P_b=17.526$ days, $x=23.448$ light-seconds, $e=2.65\times10^{-5}$, $\chi^2=188.01$, and a weighted rms residual of $5.08\,\mu$s. For a pulsar mass of $1.4\,M_\odot$ and inclination of $60^\circ$, the inferred companion mass is $0.67\,M_\odot$, favoring a CO/ONeMg white dwarf.

\end{abstract}


\section{INTRODUCTION}
Pulsars are rapidly rotating, highly magnetized neutron stars. Their extremely high densities, strong gravitational fields, and intense magnetic environments make them natural laboratories for studying extreme physical processes. Due to their remarkably stable rotational properties, the pulsed signals from pulsars serve as precise celestial clocks. Through long-term monitoring of pulse arrival times and the development of coherent timing models, pulsar timing enables accurate characterization of physical properties such as spin, astrometric position, proper motion, and binary orbital evolution. This high-precision measurement technique has yielded significant results across a broad range of fields. In particular, high-precision timing of binary pulsars provides a unique method for testing gravitational theories~\citep{1989ApJ...345..434T,2014ApJ...787...82F,2021PhRvX..11d1050K} and allows the study of general relativistic effects, including orbital decay, periastron advance, and geodetic precession~\citep{1974CRASM.279..971D,1975ApJ...199L..25B,1998ApJ...509..856K,2019ApJ...873L..15V,2023MNRAS.523.5064C,2024ApJ...966...46M,2024ApJ...964L...7Z}. Additionally, pulsar timing facilitates precise measurements of the masses of neutron stars and their companions in binary systems, providing critical observational constraints on the equation of state of high-density nuclear matter~\citep{2016ARA&A..54..401O}. Millisecond pulsars (MSPs), which exhibit extremely stable rotation and ultra-high timing precision, can be combined into pulsar timing arrays (PTAs). By monitoring the arrival-time residuals of a set of MSPs over extended periods, PTAs address fundamental issues such as neutron star rotational stability, interstellar medium perturbations, and solar-system ephemeris errors. Moreover, PTAs represent one of the most important observational tools for detecting nanohertz-frequency gravitational waves~\citep{2024ResPh..6107719V}. In recent years, several PTA collaborations—including NANOGrav, EPTA/InPTA, PPTA, and CPTA—have successively reported evidence for a nanohertz gravitational-wave background~\citep{2023ApJ...951L...8A,2023A&A...678A..50E,2023ApJ...951L...6R,2023RAA....23g5024X}. All of these applications ultimately rely on a phase-connected timing solution, in which the integer number of pulsar rotations is tracked unambiguously across the entire observing span.

With the operation of new-generation, high-sensitivity radio telescopes such as FAST and MeerKAT, along with their large-scale survey projects, the sample of known pulsars has been growing rapidly. In particular, the FAST Galactic Plane Pulsar Snapshot survey (GPPS) and the Commensal Radio Astronomy FAST Survey (CRAFTS) has discovered a large number of new pulsars, significantly expanding the sample sizes of normal pulsars, millisecond pulsars, and RRATs~\citep{2025RAA....25a4001H,2025ApJ...982..117X,2026ApJ...997..210G,2021MNRAS.508..300C,2023MNRAS.518.1672M,2024ApJ...964L...7Z,2024ApJ...975...88Z}. Surveys such as MeerKAT/TRAPUM are also continuously discovering new pulsars in environments including extragalactic systems and globular clusters, providing richer observational samples for studying pulsar populations, formation and evolution, and extreme physics~\citep{2026ApJ..1001..119S,2026MNRAS.545f2193P}.

Sparse follow-up observations often leave uncertain pulse counts across long gaps. Manual timing connects observations incrementally, refitting the spin and orbit as the counts are revised. \citet{2018MNRAS.476.4794F} automated this process by testing integer-turn assignments and comparing their timing fits. The Algorithmic Pulsar Timer (APT; \citealt{2022AJ....163...84P}) automates phase connection for isolated pulsars, using F-tests and parameter covariances to select observations, add parameters, and compare models. The Algorithmic Pulsar Timer for Binaries (APTB; \citealt{2024ApJ...964..128T}) extends this approach to binary parameters and pulse counts, using F-tests and the approximately quadratic variation of reduced $\chi^2$ near the best phase wrap.

Here, we present APOLLO-Timing (Annealed Phase--Orbit Lattice Linking across Observations), a Bayesian method for jointly determining the orbit and rotation count of a binary pulsar. The phase variation within an observing session carries information about spin evolution and orbital Doppler modulation even when the pulse count between sessions is unknown. Requiring all sessions to follow the same spin evolution and orbital light-travel-time delay combines constraints that may be inconclusive in individual observations. We group TOAs into sessions, separate the measured phase variation from an unknown integer offset for each session, and infer the offsets together with the timing parameters. Uncertainties in the common parameters correlate the session offsets. Within each orbital-cycle cell, marginalization over the local continuous-parameter uncertainty gives a lattice distribution over candidate pulse paths. We retain several paths and follow them as the added phase noise is reduced to zero. PINT~\citep{2021ApJ...911...45L} provides the physical phase model throughout the search and performs the final fits with the retained pulse numbers held fixed. These fits and the associated consistency checks constitute the certification described in Section~\ref{sec:methods}.

We test APOLLO-Timing on PSR J1912$-$1215, starting with the position, $F_0$, and $F_1$, with DM fixed and no orbital parameters supplied. We then remove sessions in reverse chronological order and compare the recovered pulse paths with the full-data solution. This comparison examines how the time span and orbital-phase coverage affect the ability to distinguish competing rotation counts. We measure the wall time from blind orbital initialization through the final timing fit, since the orbit and phase connection must be reassessed as follow-up observations accumulate. Such repeated searches are part of the timing follow-up of discoveries from large radio pulsar surveys.

The APOLLO-Timing source code is publicly available at \url{https://github.com/flagtiya/apollo_timing}.

Section~\ref{sec:methods} describes the methodological framework of APOLLO-Timing. Section~3 presents its application to PSR J1912$-$1215 and discusses the nature and possible evolutionary history of the system. Section~\ref{sec:conclusion} summarizes the main conclusions of this study.

\section{METHODS}
\label{sec:methods}

Pulsar timing involves comparing measured TOAs with predictions generated by a timing model that incorporates the pulsar's rotation, astrometric and propagation effects, and, for binary pulsars, orbital motion. TOAs are estimated by matching observed, integrated pulse profiles to a template with a defined fiducial phase. A phase-connected timing solution consistently accounts for the integer number of rotations between observations by using a single timing model throughout the entire data span.

Uncertainty in the spin and orbital parameters can leave the number of rotations across an observing gap undetermined. Several combinations of pulse counts and timing parameters may fit sparse observations. APOLLO-Timing searches these combinations jointly, using phase changes within observations and offsets between them.

After accounting for Solar-System and binary delays, the timing model expands the spin phase about a reference epoch:

\begingroup
\begin{equation}
\phi_{\rm spin}(t)=\phi_0+F_0 (t-t_0)+\frac{1}{2}F_1(t-t_0)^2+\frac{1}{6}F_2(t-t_0)^3+\cdots.
\end{equation}
\endgroup

Here $\phi_0$ is the phase at $t_0$ (PEPOCH), and $F_0$, $F_1$, and $F_2$ are the spin frequency and its first two time derivatives. Phase is measured in rotations and time differences in seconds. Let $\theta$ denote the spin, astrometric, and orbital parameters. With the pulse-profile reference set to phase zero, a phase-connected model assigns each TOA an integer pulse number $N_i$ satisfying

\begingroup
\begin{equation}
\phi(t_i;\theta) \simeq N_i.
\end{equation}
\endgroup

The difference between the model phase and assigned pulse number is

\begin{equation}
\Delta \phi_i = \phi(t_i;\theta)-N_i,
\end{equation}

or, in time units,

\begin{equation}
R_i=\frac{\Delta \phi_i}{F_0}.
\end{equation}

For a given pulse-number assignment, we fit $\theta$ by weighted least squares. APOLLO-Timing searches the allowed assignments together with the timing parameters. PINT evaluates the model phases and their parameter dependence throughout the search and performs the final fixed-count fits.

\begin{figure*}[t]
    \centering
    \includegraphics[width=\textwidth]{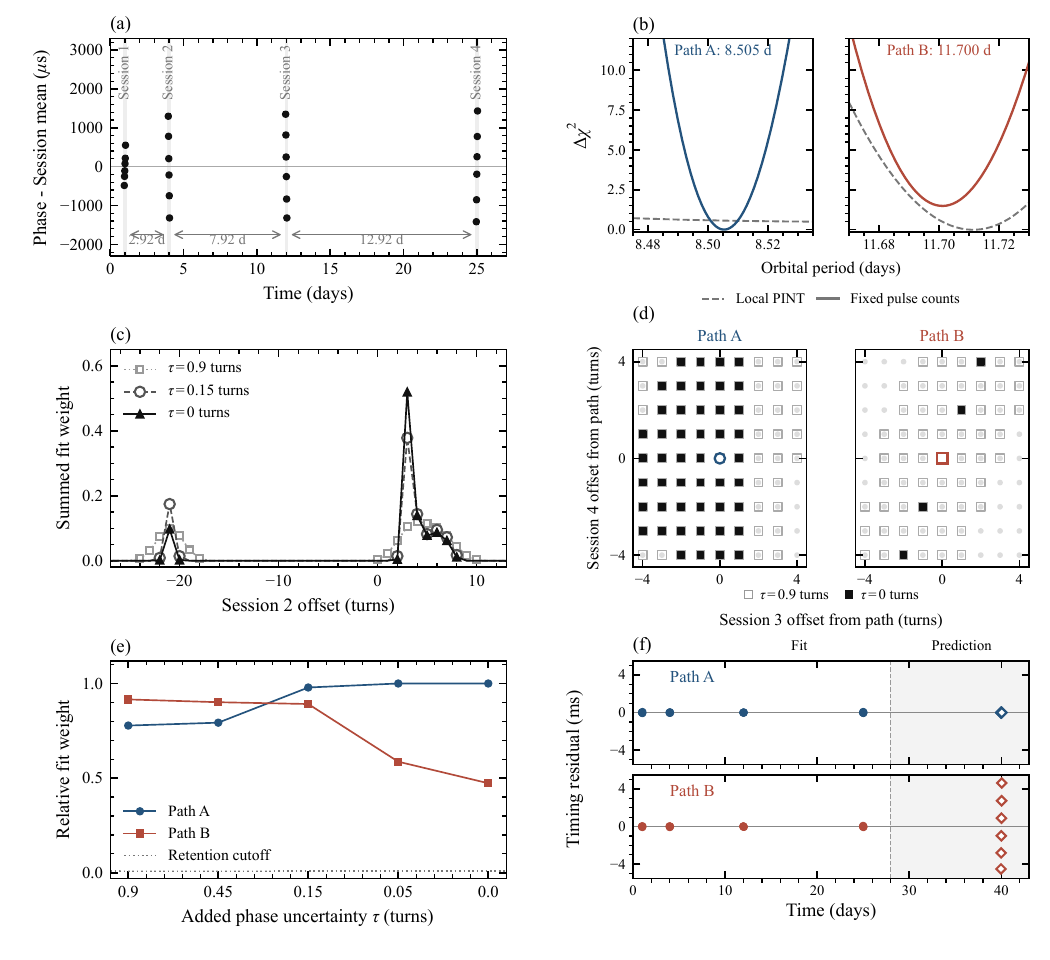}
    \caption{Phase connection of synthetic TOAs from an 8.500-day circular orbit. Each session contains six TOAs over two hours. The added phase uncertainty $\tau$, measured in turns, is reduced to zero during the search. (a) Phases after subtracting each session's mean, with arrows marking observing gaps. (b) Separate period ranges near Path A (blue) and Path B (red). Dashed curves use local PINT fits, and solid curves refit continuous parameters at fixed pulse counts. A and B share a score reference within each method. (c) Session 2 weights summed over the other offsets. (d) Integer choices near A and B, relative to Session 3/4 offsets $(-2,1)$ and $(266,545)$. Gray outlines and black squares mark retained projections at $\tau=0.9$ and $\tau=0$ turns. Session 2 is projected out. (e) Weights during annealing, with 311 paths above the final 0.01 cutoff. (f) Earlier fitted TOAs (filled circles) and independent day-40 TOAs (open diamonds, shaded region). The spin and orbital parameters remain fixed at their earlier fits, with a free mean phase for the new session. Later error bars include measurement and conditional prediction uncertainties. Both paths fit the earlier observations. The later session remains consistent with the injected Path A and gives large residuals for Path B, distinguishing the two connections.}
    \label{fig:apollo-binary-flow}
\end{figure*}

Figure~\ref{fig:apollo-binary-flow} illustrates the main steps with synthetic TOAs from a circular binary. The example combines local PINT fits with a reduced phase model for the joint pulse-count calculation. We selected a noise realization with competing orbital aliases and searched periods of 6--14 days and offsets of $\pm1000$ turns. The plotted weights come from profile fits on this finite grid and do not include marginalization over timing-parameter uncertainty. 

\subsection{Phase Information within Observations}

TOAs within an observation usually constrain relative pulse counts more tightly than TOAs separated by a long gap. We group TOAs by observing epoch, combining closely spaced scans into a session using a time-gap threshold. A count error across a gap shifts every pulse number in a session by the same integer, leaving their differences unchanged. We write the unknowns as

\begin{equation}
\{\theta,\mathbf m,\boldsymbol{\nu}\},
\end{equation}

where $\theta$ contains the fitted timing parameters, $\mathbf m$ the within-session count corrections, and $\boldsymbol{\nu}$ the whole-session integer offsets. For $S$ sessions,

\begingroup
\begin{equation}
\boldsymbol{\nu}=(n_1,n_2,\ldots,n_{S-1})^{\mathsf T}.
\end{equation}
\endgroup

Setting $n_0=0$ fixes the arbitrary count origin: a common integer shift leaves all pulse-number differences unchanged. We call a complete assignment a pulse-number path. For TOA $i$ in session $s$,

\begingroup
\begin{equation}
K_{s,i}=K_{s,i}^{(0)}+m_{s,i}+n_s.
\end{equation}
\endgroup

The input model supplies the reference count $K_{s,i}^{(0)}$, to which we add the within-session correction $m_{s,i}$ and the whole-session offset $n_s$. Here $K_{s,i}$ is the session-indexed form of $N_i$. We retain alternative within-session counts when needed.

Within-session phase changes constrain the model before $n_s$ is known. Let $\boldsymbol{\Phi}_s(\theta)$ contain the model phases. For a trial within-session count assignment, the residual vector is

\begin{equation}
\mathbf r_s=\boldsymbol{\Phi}_s(\theta)-\mathbf K_s^{(0)}-\mathbf m_s.
\end{equation}

For Gaussian TOA errors, the inverse phase-error covariance $\mathbf W_s$ gives the measurement weights. We separate the weighted mean offset $a_s$ from the remaining residuals $\mathbf q_s$:

\begingroup
\begin{equation}
\begin{aligned}
a_s&=\frac{\mathbf 1_s^{\mathsf T}\mathbf W_s\mathbf r_s}
{\mathbf 1_s^{\mathsf T}\mathbf W_s\mathbf 1_s},\\
\mathbf q_s&=\mathbf W_s^{1/2}\mathbf P_s\mathbf r_s.
\end{aligned}
\end{equation}
\endgroup

Here $\mathbf 1_s$ and $\mathbf I_s$ are the all-ones vector and identity matrix, with dimensions set by the number of TOAs in session $s$. The projection

\begingroup
\begin{equation}
\mathbf P_s=\mathbf I_s-\mathbf 1_s
\left(\mathbf 1_s^{\mathsf T}\mathbf W_s\mathbf 1_s\right)^{-1}
\mathbf 1_s^{\mathsf T}\mathbf W_s
\end{equation}
\endgroup

removes the common phase offset, as a separate JUMP does for a group of TOAs in DRACULA or APTB~\citep{2018MNRAS.476.4794F,2024ApJ...964..128T}. The remaining slope and curvature constrain the timing model. We obtain the offset from the weighted mean and scale the residuals by $\mathbf W_s^{1/2}$, so their squared sum accounts for the measurement uncertainties.

To connect sessions, we compare their mean offsets with session 0: $\mathbf d=(a_1-a_0,\ldots,a_{S-1}-a_0)^{\mathsf T}$. With $v_s=(\mathbf 1_s^{\mathsf T}\mathbf W_s\mathbf 1_s)^{-1}$, the covariance of these differences is

\begin{equation}
\mathbf V=\operatorname{diag}(v_1,\ldots,v_{S-1})+v_0\mathbf 1\mathbf 1^{\mathsf T},
\end{equation}

where $\mathbf 1$ has length $S-1$. The shared uncertain reference produces the off-diagonal entries. A trial model must fit both $\mathbf q_s$ and the relative mean offsets $\mathbf d$, with covariance $\mathbf V$.

\subsection{Trial Orbits and the Orbital-Period Search}

When no initial orbital solution is available, or only a subset of its parameters is known, we estimate the missing orbital parameters from phase changes within observing sessions. The resulting trial orbits are evaluated with the selected PINT binary model.

The orbital period $P_b$ sets the rate of phase advance, and the projected semimajor axis $x=a_{\rm p}\sin i/c$ sets the Roemer-delay amplitude. Here $a_{\rm p}$ is the pulsar's semimajor axis about the binary barycentre, $i$ is inclination, and $e$ is eccentricity. The parameter file denotes $x$ by $A_1$, in light-travel time. For nearly circular orbits, ELL1 uses $T_{\rm ASC}$ (TASC), $\epsilon_1=e\sin\omega$ (EPS1), and $\epsilon_2=e\cos\omega$ (EPS2) to describe orbital phase and eccentricity corrections. These coordinates avoid the poorly determined periastron epoch $T_0$ and longitude $\omega$ at small eccentricity. ELL1H and ELL1k use the same coordinates. BT, DD, DDS, DDGR, DDH, and DDK use $T_0,e,\omega$. BT-piecewise also fits offsets between orbital segments.

For a trial orbital period, an initial parameter set for an ELL1 model may contain only

\begingroup
\begin{equation}
\theta_c=(\alpha,\delta,F_0,F_1,t_{\rm epoch},P_b).
\end{equation}
\endgroup

Here $\alpha,\delta$ are right ascension and declination, and $t_{\rm epoch}$ is the spin reference epoch. We calculate phases without the binary delay and select a densely sampled observing session. Its closely spaced TOAs trace the residual phase during the observation, constraining its slope and any measurable curvature. A circular orbit contributes sine and cosine terms, whose changes we fit relative to the first TOA. Let $w_i$ be the phase modulo one rotation and $t_0$ the session's first TOA. For $\lambda_i=2\pi(t_i-t_0)/P_b$,

\begin{equation}
\mathbf D_i=(\sin\lambda_i-\sin\lambda_0,\;
\cos\lambda_i-\cos\lambda_0).
\end{equation}

The coefficients $\mathbf c=(u,v)^{\mathsf T}$ give the orbital phase term. We fit them jointly with the within-session count corrections $\mathbf k$, weighted by the phase errors $\sigma_{\phi,i}$. An upper bound $x_{\max}$ restricts the amplitude to $|F_0|x_{\max}$:

\begin{equation}
\begin{aligned}
p(\mathbf c,\mathbf k\mid\mathbf w,P_b)
&\propto \mathbf 1\!\left(\|\mathbf c\|_2\le |F_0|x_{\max}\right)\\
&\quad\times\prod_i
\exp\!\left[-\frac{(w_i+\mathbf D_i\mathbf c-k_i)^2}
{2\sigma_{\phi,i}^2}\right].
\end{aligned}
\end{equation}

The indicator $\mathbf 1(\cdot)$ equals one when the bound holds and zero otherwise. We increase $x_{\max}$ through a specified sequence, then refine candidates with a second harmonic and a local $F_0$ correction. The coefficients are converted to $A_1,T_{\rm ASC},\mathrm{EPS1},\mathrm{EPS2}$ and checked with the PINT ELL1 delay. These trial orbits seed the phase-connection search.

If an initial solution provides only some of the orbital parameters, we estimate the missing values in the appropriate Keplerian coordinates:

\begingroup
\begin{equation}
q_{\rm Kep}=\begin{cases}
(A_1,T_{\rm ASC},\epsilon_1,\epsilon_2),&\text{ELL1 models}\\
(A_1,T_0,e,\omega),&\text{eccentric models}
\end{cases}.
\end{equation}
\endgroup

For eccentric orbits, we search a finite grid in eccentricity, orbital phase, and periastron longitude. Compatible within-session counts are retained, and the specified PINT model refines the trial orbits. Multiple starts reduce dependence on a single coarse fit. Initialization supplies the basic orbit without adding $F_2$, $\dot P_b$, $\dot x$, $\dot\epsilon_1$, or $\dot\epsilon_2$ automatically. Complete initial parameter sets retain their specified fit flags.

We scale parameter changes by their phase effects to search spin, position, and orbit together. PINT evaluates the physical timing model at each trial. Let $T_{\rm span}$ be the observing span and $t_r$ the search reference epoch. With the light-travel time across one astronomical unit, $t_{\rm AU}=499.004783836\,{\rm s}$, the spin and position coordinates are

\begin{equation}
\begin{aligned}
{F_r}&=F_0+F_1(t_r-\mathrm{PEPOCH})_{\rm s},\\
\kappa_{\rm spin}&=\frac{1}{2}F_1(T_{\rm span})_{\rm s}^{2},\\
\xi_E&=F_r t_{\rm AU}\cos\delta_0\,\Delta\alpha,\\
\xi_N&=F_r t_{\rm AU}\,\Delta\delta.
\end{aligned}
\end{equation}

To the order shown, $F_r$ is the frequency at $t_r$ and $\kappa_{\rm spin}$ is the quadratic phase scale across the span. The subscript s denotes seconds. Position offsets $\Delta\alpha,\Delta\delta$ from $(\alpha_0,\delta_0)$ are in radians, and $\xi_E,\xi_N$ give their phase scales through the annual Roemer delay. The time dependence also depends on source direction and Earth's position. For ecliptic coordinates, the longitude offset carries a factor $\cos\beta_0$, where $\beta_0$ is the input latitude.

Sparse observations may leave the number of elapsed orbits uncertain. We use the orbital cycle count $C_{\rm orb}$ across the data and the phase at $t_r$:

\begingroup
\begin{equation}
\begin{aligned}
C_{\rm orb}&=\frac{T_{\rm span}}{P_b},\\
\phi_r&=\operatorname{wrap}\!\left[2\pi\frac{t_r-T_{\rm orb}}{P_b}\right].
\end{aligned}
\end{equation}
\endgroup

Here $T_{\rm orb}$ is $T_{\rm ASC}$ for ELL1, ELL1H, and ELL1k and $T_0$ for eccentric models, and $\operatorname{wrap}$ places phase in one $2\pi$ interval. Changing $C_{\rm orb}$ by one adds or removes one orbit over $T_{\rm span}$. Sparse observations can admit several such aliases. We search finite intervals in $C_{\rm orb}$, termed orbital-cycle cells, with a specified maximum width.

The other orbital coordinates describe delay amplitude and shape. For ELL1, $\mathbf a_{\rm ELL1}=(F_rA_1,F_rA_1\mathrm{EPS1},F_rA_1\mathrm{EPS2})$ gives the main term and eccentricity corrections in phase units. For eccentric models, $\mathbf a_{\rm ecc}$ contains $\arcsin\sqrt e$ and sine--cosine coordinates for amplitude and periastron longitude at $t_r$. A fitted OMDOT is scaled to the periastron angle accumulated over $T_{\rm span}$. We collect these coordinates as

\begingroup
\begin{equation}
q_M=\begin{cases}
(C_{\rm orb},\phi_r,F_r,\mathbf a_{\rm ELL1},\ldots),
&\text{ELL1 models}\\
(C_{\rm orb},\phi_r,F_r,\mathbf a_{\rm ecc},\ldots),
&\text{eccentric models}
\end{cases}.
\end{equation}
\endgroup

For models with orbital segments, we append the fitted segment offsets $\mathbf b_{\rm pw}$ to the eccentric-model coordinates. The omitted entries are other free timing parameters. Scaling by the allowed ranges gives the dimensionless vector $\mathbf z$, which we convert to physical parameters for each PINT evaluation. The allowed parameters and integer offsets define the search range, or support. The priors specify the relative weights within these bounds. The numbers of starts and retained paths are finite.

\subsection{Joint Fits of Timing Parameters and Pulse Numbers}

For each trial orbit, we seek pulse counts that connect all sessions with one timing model. Let $\mathbf y$ collect the within-session residuals $\mathbf q_s$. Near a trial model, the first-order response of $\mathbf y$ and the inter-session offsets to $\delta\mathbf z$ is

\begin{equation}
\begin{aligned}
\mathbf y(\mathbf z+\delta\mathbf z)&\simeq\mathbf y_0+\mathbf A\delta\mathbf z,\\
\mathbf d(\mathbf z+\delta\mathbf z)&\simeq\mathbf d_0+\mathbf B\delta\mathbf z.
\end{aligned}
\end{equation}

Each column of $\mathbf A$ or $\mathbf B$ gives the phase response to a small parameter change, evaluated with PINT. This linear approximation allows timing uncertainty and pulse counts to be compared jointly.

A connection must fit the within-session phase changes and the differences between session means. Let $D$ denote the TOAs and $c$ the orbital interval. Combining the Gaussian likelihood and priors, with added phase uncertainty $\tau$ (Section 2.4), gives

\begin{equation}
\begin{aligned}
p(\mathbf z,\mathbf m,\boldsymbol{\nu},c\mid D,\tau)
&\propto p(c)\,p(\mathbf z\mid c)\\
&\quad\times p(\mathbf m,\boldsymbol{\nu}\mid c)
\exp\!\left(-\frac{1}{2}\mathcal Q_\tau\right),\\
\mathcal Q_\tau
&=\sum_s\|\mathbf q_s(\mathbf z,\mathbf m_s)\|^2
+\mathbf e^{\mathsf T}\mathbf V_\tau^{-1}\mathbf e,\\
\mathbf e
&=\mathbf d(\mathbf z,\mathbf m)-\boldsymbol{\nu}.
\end{aligned}
\end{equation}

The weighted squared phase differences $\mathcal Q_\tau$ play the role of a timing $\chi^2$. The two terms measure the within-session residuals and the differences between predicted and integer session offsets, including correlated errors. The priors weight the orbital interval, timing parameters, and counts. For an isolated pulsar, $c$ is fixed because there is no orbital-period search.

Within each orbital interval, starts cover a finite grid in cycle count and phase and include the interval centre. We also use any supplied parameter-prior component means that lie in the interval. Bounded PINT fits refine these starts. Around each resulting model, we adopt the local Gaussian prior $\delta\mathbf z\sim\mathcal N(\mathbf 0,\boldsymbol{\Lambda})$, whose covariance $\boldsymbol{\Lambda}$ sets the parameter widths and correlations. The within-session data give

\begin{equation}
\begin{aligned}
\boldsymbol{\Sigma}_y&=\left(\boldsymbol{\Lambda}^{-1}+\mathbf A^{\mathsf T}\mathbf A\right)^{-1},\\
\boldsymbol{\mu}_y&=-\boldsymbol{\Sigma}_y\mathbf A^{\mathsf T}\mathbf y_0.
\end{aligned}
\end{equation}

The within-session fit leaves uncertainty in the timing parameters. Marginalizing over their local distribution gives the predicted mean $\mathbf c$ and covariance $\mathbf S_\tau$ of the session offsets:

\begin{equation}
\begin{aligned}
\mathbf c&=\mathbf d_0+\mathbf B\boldsymbol{\mu}_y,\\
\mathbf S_\tau&=\mathbf V_\tau+\mathbf B\boldsymbol{\Sigma}_y\mathbf B^{\mathsf T}.
\end{aligned}
\end{equation}

The contribution $\mathbf B\boldsymbol{\Sigma}_y\mathbf B^{\mathsf T}$ propagates the shared timing-parameter uncertainty and correlates the offsets. Measurement errors and added search uncertainty enter through $\mathbf V_\tau$, with $\mathbf V_0=\mathbf V$. In $\mathbf S_\tau$, diagonal entries are variances and off-diagonal entries are covariances. The offset vector $\mathbf c$ is distinct from the initialization's harmonic coefficients.

At fixed orbital interval and within-session assignment, equal priors on allowed integer paths give the local Gaussian distribution

\begin{equation}
p(\boldsymbol{\nu}\mid\mathbf y,\mathbf d,\tau)
\propto\exp\!\left[-\frac{1}{2}
(\boldsymbol{\nu}-\mathbf c)^{\mathsf T}\mathbf S_\tau^{-1}
(\boldsymbol{\nu}-\mathbf c)\right].
\end{equation}

Each integer vector assigns a set of session pulse counts. Increasing one component by one adds a rotation to that session and moves to a neighboring lattice point. We evaluate complete vectors with the covariance of the predicted offsets, accounting for correlations from the shared timing parameters. Their uncertainty-weighted distance from the prediction is

\begin{equation}
Q_{k,\tau}=(\boldsymbol{\nu}_k-\mathbf c)^{\mathsf T}
\mathbf S_\tau^{-1}(\boldsymbol{\nu}_k-\mathbf c).
\end{equation}

For path $k$, $Q_{k,\tau}$ is the squared Mahalanobis distance, including offset uncertainties and correlations. Smaller values give larger local weights. We enumerate a finite set of paths at each start in every orbital interval. To assess contributions from omitted paths near a trial model, we sum the first $K$ weights:

\begingroup
\begin{equation}
Z_{\mathbb Z,K}(\tau)=\frac{1}{(2\pi)^{(S-1)/2}|\mathbf S_\tau|^{1/2}}
\sum_{k=1}^{K}\exp\!\left(-\frac{Q_{k,\tau}}{2}\right).
\end{equation}
\endgroup

This Gaussian-normalized sum is the truncated lattice mass. At fixed $\tau$, we increase $K$: continued growth indicates omitted local weight, while stabilization means additional paths contribute little near this model. The check cannot exclude distant timing solutions or orbital aliases. Section~\ref{sec:path-stability} reports the convergence diagnostics.

\subsection{Refining Candidate Connections}

An inaccurate trial spin or orbit can make a viable pulse assignment fit poorly. To avoid rejecting it before refitting, we initially allow extra uncertainty in session offsets and reduce it through $\tau_0>\tau_1>\cdots>\tau_L=0$, in rotations. At each stage,

\begingroup
\begin{equation}
\mathbf V_\tau=\mathbf V+\tau^2
(\mathbf I_{S-1}+\mathbf 1_{S-1}\mathbf 1_{S-1}^{\mathsf T}).
\end{equation}
\endgroup

Here $\mathbf I_{S-1}$ is the identity matrix and $\mathbf 1_{S-1}$ the all-ones vector. The added term represents independent session phase variances $\tau^2$, whose differences from the shared reference have off-diagonal covariances. This search uncertainty leaves the within-session weights unchanged. At $\tau=0$, only the measurement covariance $\mathbf V$ remains.

At each stage, bounded fits update the timing parameters and integer enumeration updates the path weights. This annealing procedure carries a finite set of the highest-weight paths forward as the added uncertainty decreases. It uses deterministic fits and enumeration without Markov chain Monte Carlo (MCMC) sampling. Paths consistent at $\tau=0$ proceed to full-model fits.

\subsection{Fits with the Full Timing Model}

For each retained path, we reconstruct $K^{\rm cert}_{s,i}=K^{(0)}_{s,i}+m_{s,i}+n_s$ and save the pulse numbers in the timing file. Starting from the $\tau=0$ parameters, PINT fits the full model by weighted least squares with pulse-number tracking enabled. All pulse numbers remain fixed.

Fit flags come from the input and initialization. At fixed pulse numbers, we compare binary models and additional timing terms using the Bayesian information criterion (BIC), their joint significance relative to the simpler model, and parameter bounds. Adding $F_2$, secular orbital derivatives, or post-Keplerian terms requires a fit improvement large enough to offset the BIC penalty.

For each path, we record parameters, residuals, $\chi^2$, reduced $\chi^2$ (per degree of freedom), and weighted rms. Reloading \texttt{solution.par} and \texttt{solution.tim} into PINT must reproduce the fit. Acceptance requires parameters within bounds, reduced $\chi^2<20$ before and after reloading, and $|\Delta\chi^2|\leq0.01$. The separate target criterion is reduced $\chi^2<10$. These PINT certification checks assess fit quality and file consistency. They cannot independently establish the pulse count.

\subsection{Solution Selection and Search Limits}

\label{sec:solution-selection}

Among accepted paths, we report the one with the smallest fixed-count PINT $\chi^2$ for the same model and TOA errors. Approximate path weights guide search retention, while the final comparison uses the full timing fits. Fit quality, accepted-path counts, and search limits are reported separately.

Coverage within each orbital interval is limited by the finite starting grid, integer enumeration, and path retention. Connections may remain undiscovered even when the local weight sum converges. We record the orbital intervals, candidate limits, phase-uncertainty schedule, and fit diagnostics. Section~\ref{sec:path-stability} tests sensitivity to selected numerical settings and the initial spin model.

The 24 scans form 23 sessions because the first day's two scans are grouped together. We test shorter spans by successively removing the latest session, retaining all TOAs in the first 23 down to the first 12 sessions. After each search, we compare counts with the full-data reference, allowing a common integer shift. Table~\ref{tab:session-recovery} reports accepted paths, reference matches, and original wall times. Section~\ref{sec:path-stability} gives the initialization and search-setting tests.

\section{RESULTS}
We applied APOLLO-Timing to the newly discovered binary pulsar PSR J1912$-$1215, for which no phase-connected timing solution has previously been published. We describe the blind search and the dependence of phase recovery on session coverage below, followed by the implications of the timing solution for the companion and the possible evolutionary history of the system.

\subsection{Application of APOLLO-Timing to PSR J1912$-$1215}
\label{sec:RESULTS}

\begin{figure}
	\includegraphics[width=\linewidth]{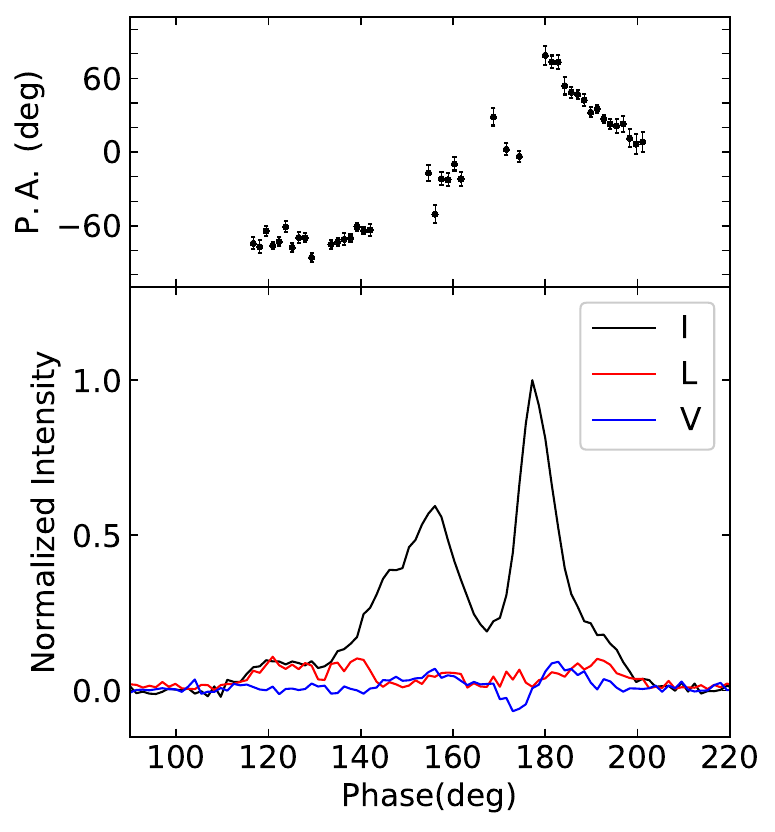}
    \caption{Integrated pulse profiles for PSR J1912$-$1215. The black line in the lower panel of each subplot represents the total intensity, and the red line and blue line represent the linear polarization intensity and circular polarization intensity respectively. From the plot, it can be seen that the peak intensity is normalized to 1.}
    \label{pola}
\end{figure}

PSR J1912$-$1215 was discovered by the CRAFTS survey~\citep{Li2018CRAFTS}. The FAST observations used here span 1438.07 days and include 24 observing scans, which form 23 observing sessions under the APOLLO-Timing session definition. Each FAST scan lasted 5 minutes and recorded full Stokes parameters with a sampling time of 49.152 $\mu s$. Additionally, to calibrate the polarization of this pulsar, a 40-second noise-diode injection was performed at the start of each observing scan. Figure~\ref{pola} shows the polarization profile of PSR J1912$-$1215.

\begin{figure}
	\includegraphics[width=\linewidth]{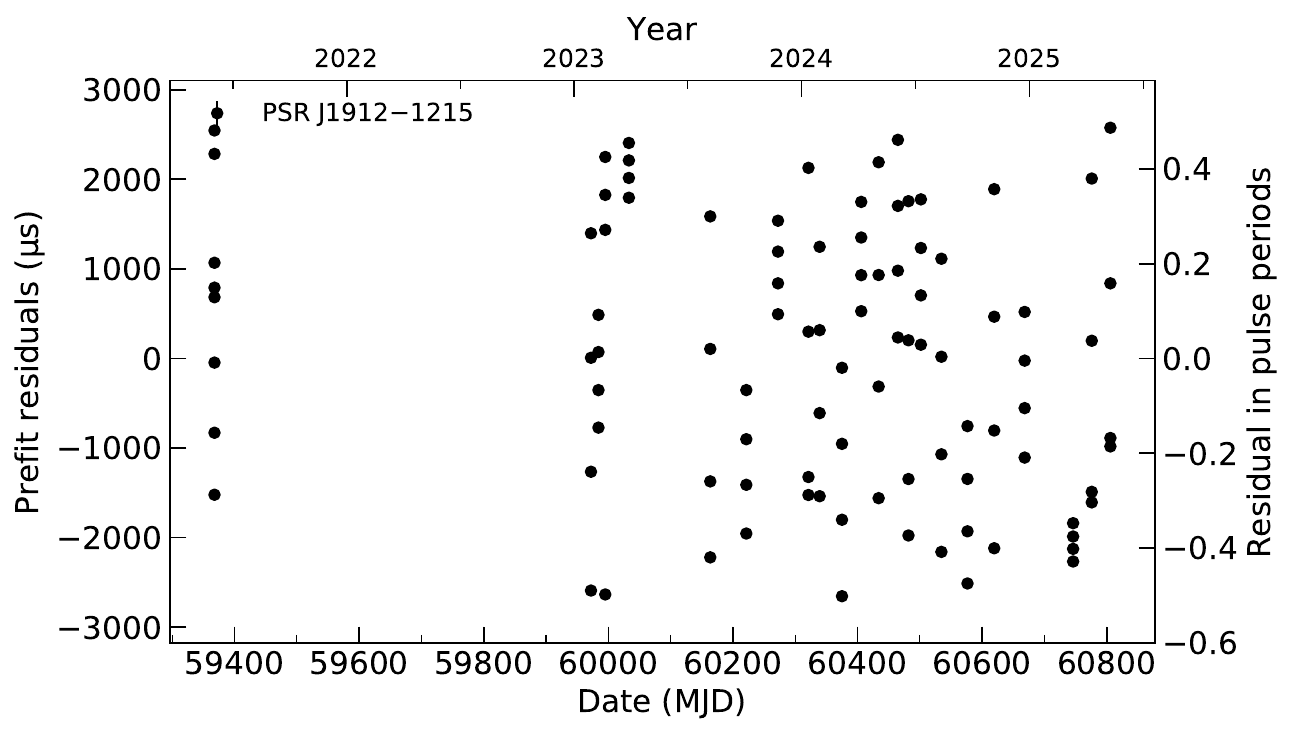}
    \caption{Timing residuals obtained from the preliminary timing solution for PSR J1912$-$1215.}
    \label{preliminary-timing}
\end{figure}

Initial follow-up observations showed significant epoch-to-epoch variations in the apparent spin period of PSR J1912$-$1215 and its derivative, consistent with radial-velocity and acceleration changes in a binary orbit. We fitted a preliminary Keplerian model with the Java-based tool newfitorbit\footnote{\href{https://github.com/SixByNine/newfitorbit}{\nolinkurl{github.com/SixByNine/newfitorbit}}}, estimating the orbital period $P_b$, periastron epoch $T_0$, projected semimajor axis $x$, and spin-frequency derivative $F_1$. The residuals from this initial timing solution (Figure~\ref{preliminary-timing}) show that it does not phase-connect all observations.

\begin{table}
\begin{threeparttable}
\footnotesize
\setlength{\tabcolsep}{0pt}
\caption{Best-ﬁtting Timing Parameters for PSR J1912$-$1215.} \label{tab:t1}
\begin{tabular*}{\linewidth}{@{\extracolsep{\fill}}ll}
\hline
Pulsar name & PSR J1912$-$1215 \\
\hline\hline
\multicolumn{2}{l}{\textbf{Measured Parameters}} \\
R.A., $\alpha$ (hh:mm:ss) & 19:12:34.49733(2) \\
Decl., $\delta$ (dd:mm:ss) & $-$12:15:21.980(2) \\
Pulse frequency, $\nu$ (s$^{-1}$) & 189.047239445402(8) \\
Frequency derivative, $\dot{\nu}$ (s$^{-2}$) &$-$3.024(2)$\times 10^{-16}$ \\
Dispersion measure, DM (pc cm$^{-3}$) & 39.495(1) \\
Faraday rotation measure, RM (rad m$^{-2}$) & $-$31(26) \\
Orbital period, $P_b$ (days) & 17.526125680(7)\\
Epoch of ascending node, TASC (MJD) & 60606.87637480(14) \\
Projected semimajor axis, $x$ (lt-s) & 23.4476454(6) \\
EPS1    & $-$2.558(6)$\times10^{-5}$ \\
EPS2    & 6.76(6)$\times10^{-6}$  \\
Binary model & ELL1 \\
\hline
\multicolumn{2}{l}{\textbf{Fitting Parameters}} \\
\hline
First TOA (MJD) &  59367.8\\
Last TOA (MJD) &  60805.9\\
Timing epoch (MJD) & 60604.0 \\
Data span (yr) & 3.94 \\
Number of TOAs & 96 \\
Solar system ephemeris model & DE421 \\
PINT $\chi^2$ & 188.01 \\
Reduced $\chi^2$ & 2.19 \\
Weighted RMS residual ($\mu s$) & 5.08 \\
APOLLO wall time (s) & 914.0 \\
Maximum RSS (MiB) & 255 \\
UNITS & TDB \\
\hline
\multicolumn{2}{l}{\textbf{Derived Parameters}} \\
\hline
$P$ (s) & 0.0052896831656132(2) \\
$\dot{P}$ (s s$^{-1}$) & 8.462(5)$\times 10^{-21}$ \\
Galactic longitude, $l$ ($^\circ$)  &  24.290\\
Galactic latitude, $b$ ($^\circ$)& $-$10.189 \\
DM distance, $d$ (kpc) \\
\multicolumn{1}{r}{YMW16\quad\mbox{}}  & 1.27\\
\multicolumn{1}{r}{NE2025\quad\mbox{}} & 1.67\\
Characteristic age, $\tau_c$ (Myr) & 9.90 $\times 10^3$ \\ 
Surface magnetic field, $B_s$ (G) & 2.14$\times 10^{8}$ \\
Spin-down luminosity, $\dot{E}$ (erg s$^{-1}$) &  2.26$\times 10^{33}$\\
\hline
\end{tabular*}
\end{threeparttable}
    \begin{tablenotes}
\item Note: The timing solution and resource measurements are from a repeat blind run with PEPOCH set to MJD 60604. Table~\ref{tab:session-recovery} reports the original session-prefix runs. The DM was determined by dividing the full observing band into two sub-bands of equal bandwidth but different central frequencies, and then fitting these sub-bands. The timing fit held DM fixed at $39.48\,\mathrm{pc\,cm^{-3}}$. The rotation measure (RM) was measured using the rmfit tool from the PSRCHIVE software package~\citep{2004PASA...21..302H}.
\end{tablenotes}
\end{table}

For the blind APOLLO-Timing calculation, we retained the position, $F_0$, and preliminary $F_1$, fixed DM, and removed the binary component and orbital parameters. The search obtained its starting orbits from the within-session phase changes before connecting the observations. We repeated this complete blind calculation with PEPOCH set to MJD 60604 for the solution reported in Table~\ref{tab:t1}. It recovered the same relative pulse counts as the original full-data run used for the chronological-prefix experiment.

\begin{figure*}
	\includegraphics[width=\linewidth]{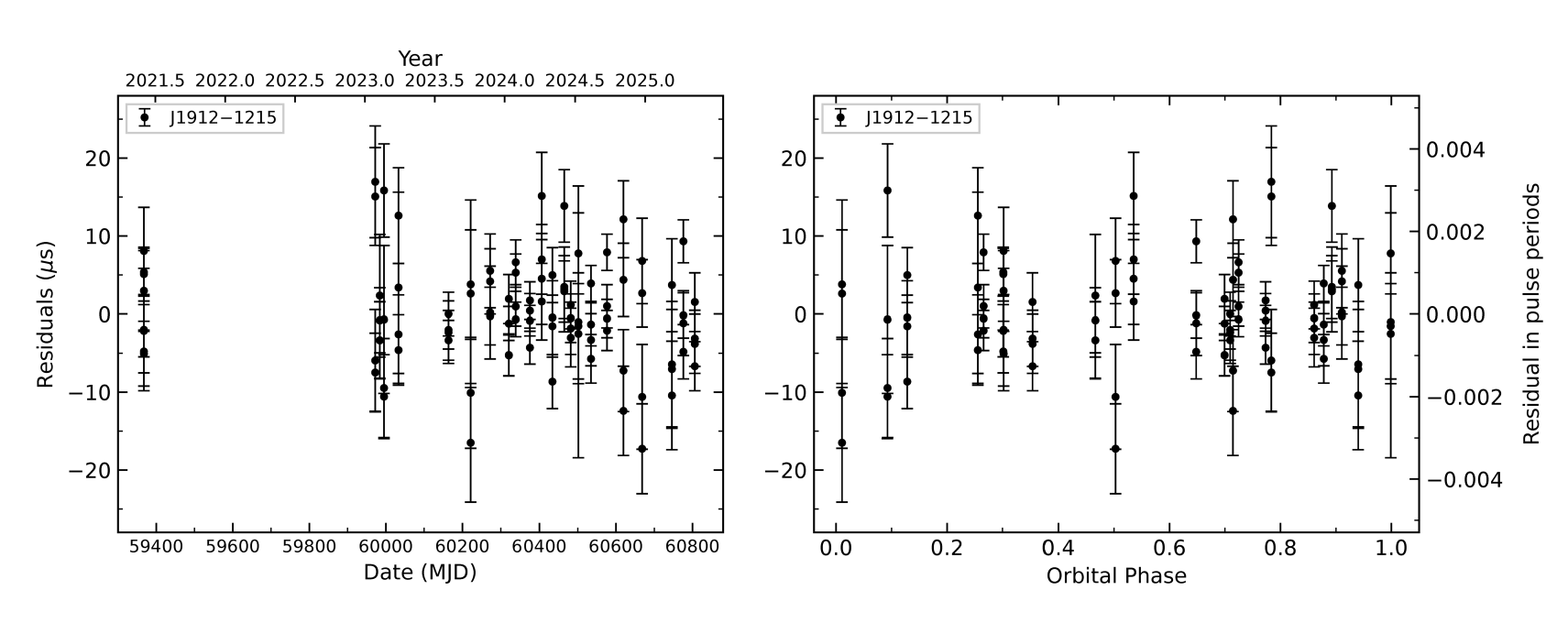}
    \caption{The post-fit residuals for PSR J1912$-$1215. The left and right panels show the residuals vs. MJD and orbital phase, respectively.}
    \label{timing}
\end{figure*}

APOLLO-Timing searched the orbital-period intervals and correlated session offsets. One candidate reached zero added uncertainty and passed the final PINT fit. The solution (Table~\ref{tab:t1}) has $\chi^2=188.01$, reduced $\chi^2=2.19$, and a weighted rms residual of $5.08\,\mu$s with the reported TOA errors. Figure~\ref{timing} shows the residuals against time and orbital phase. No systematic orbital-phase trend, including a Shapiro-delay signature, is apparent. The YMW16~\citep{2017ApJ...835...29Y} and NE2025~\citep{2026ApJ..1002....3O} electron-density models give DM-based distances of 1.27 and 1.67 kpc.

Table~\ref{tab:session-recovery} follows recovery as the latest sessions are removed. It counts accepted paths and matches to the full-data reference separately, allowing a common integer shift. Fits that fail the reduced-$\chi^2$ or reload checks are excluded from the accepted count. The 15-, 17-, 18-, and 20- to 23-session runs retained only the reference. The 16-session run also retained one alternative. The second 19-session fit had different counts but failed the checks. At 12 to 14 sessions, accepted paths differed from the reference. Each removal changes the baseline and orbital-phase coverage, as well as the finite candidate set retained by the search.

\begin{table*}
\centering
\caption{Blind recovery for chronological session subsets of PSR J1912$-$1215, using PEPOCH = MJD 59368. ``Accepted paths'' counts distinct assignments covering all input TOAs and passing the fit-quality and reload checks. ``Reference'' counts matches to the full-data assignment. A common integer shift is allowed in both counts. The second 19-session fit failed the checks and is excluded. Wall times were measured on the same single-threaded host.}
\label{tab:session-recovery}
\begin{tabular}{r r l r r r}
\hline
Sessions & TOAs & Outcome & Accepted paths & Reference & Wall time (s) \\
\hline
23 & 96 & Reference recovered & 1 & 1 & 922.9 \\
22 & 92 & Reference recovered & 1 & 1 & 892.4 \\
21 & 88 & Reference recovered & 1 & 1 & 834.2 \\
20 & 84 & Reference recovered & 1 & 1 & 787.8 \\
19 & 80 & Reference recovered & 1 & 1 & 771.1 \\
18 & 76 & Reference recovered & 1 & 1 & 757.9 \\
17 & 72 & Reference recovered & 1 & 1 & 660.7 \\
16 & 68 & Reference plus one other path & 2 & 1 & 681.2 \\
15 & 64 & Reference recovered & 1 & 1 & 668.6 \\
14 & 60 & Different accepted path & 2 & 0 & 650.4 \\
13 & 56 & Different accepted path & 1 & 0 & 597.9 \\
12 & 52 & Different accepted path & 2 & 0 & 620.8 \\
\hline
\end{tabular}
\end{table*}

\subsubsection{Search Stability and Alternative Pulse Paths}
\label{sec:path-stability}

To examine the extra path at 16 sessions, we retained its pulse numbers for the first 15 sessions (64 TOAs) and refitted the timing parameters. The result had $\chi^2=153.60137$, reduced $\chi^2=2.84447$, and reload difference $|\Delta\chi^2|=2.38\times10^{-5}$. All parameters stayed within the original 15-session bounds, and the path passed the same acceptance checks while remaining distinct from the reference after a common shift. Thus an acceptable alternative existed even though the original 15-session search retained only the reference. This fixed-path test leaves the blind-search counts and times in Table~\ref{tab:session-recovery} unchanged.

Fifteen additional runs of the 17- to 19-session subsets tested a wider phase grid, larger path-retention and enumeration limits, a zero starting $F_1$, and a wider $F_1$ range. All selected the reference as their best fit. The coarse accumulated path weight did not converge in any run, leaving search completeness unverified. These tests assess the stability of the preferred path under the tested settings. Independent fixed-path posterior weights and evidence ratios were not calculated.

Starting at $F_1=0$ recovered the reference at 17, 18, and 19 sessions. With a wider $F_1$ range, the 18-session search also found an accepted alternative with $\chi^2=498.39$, reduced $\chi^2=7.55$, and reload $\Delta\chi^2=0.0027$. Its $F_1=+2.39\times10^{-13}\,{\rm Hz\,s^{-1}}$ and eight other fitted parameters lie within the original bounds. It is poorer than the reference by $\Delta\chi^2=373.87$. For the same Gaussian error model, its maximum likelihood is $6.5\times10^{-82}$ times that of the reference. This ratio compares the maximum likelihoods without integrating over timing parameters or including prior path probabilities. The acceptance thresholds admit this much poorer fit in the 18-session subset. The full 96-TOA result is unchanged.

Parameter bounds set the coordinate scales and local Gaussian prior covariance, so widening them changes both the search and the effective prior. We imposed neither $F_1<0$ nor a stellar-companion-mass restriction on $P_b$ and $x$. Such astrophysical priors require justification and sensitivity tests, especially over a short timing baseline. They should not be chosen merely to remove alternatives. Here, fit quality favors the reference over the alternative found with the wider $F_1$ range.

\subsection{Origin of the system}
\label{sec:system origin}

Based on the timing results we have obtained, we can infer the mass range of the companion star using the mass function.
\begin{equation}
\frac{f(m_{\mathrm{p}},m_{\mathrm{c}})}{M_\odot}
=\frac{(m_{\mathrm{c}}\sin i)^3}{M^2 M_\odot}
=\frac{4\pi^2 x^3}{T_\odot P_b^2},
 	\label{eq:quadratic}
\end{equation}
where $M=m_{\rm p}+m_{\rm c}$ is the total mass of the binary system, with $m_{\rm p}$ and $m_{\rm c}$ denoting the pulsar and companion masses, respectively. The constant $T_\odot \equiv G M_\odot/c^3 = 4.925490947~\mu{\rm s}$ represents the solar mass expressed in time units, where $G$ is the gravitational constant and $c$ is the speed of light. Here $f$, $m_{\rm p}$, $m_{\rm c}$, and $M$ are dimensional masses. The quantities $x$ and $P_b$ must be expressed in seconds when $T_\odot$ is inserted in seconds. Assuming a canonical pulsar mass of $m_{\rm p}=1.4~M_\odot$ and an orbital inclination angle of 
$i=60^\circ$, we obtain a companion mass of $m_{\rm c}=0.67~M_\odot$ for PSR J1912$-$1215 (the corresponding minimum mass at $i=90^\circ$ is $0.56~M_\odot$). The inferred intermediate companion mass falls within the white dwarf mass gap identified by~\cite{2023MNRAS.520.1789S}, specifically the transitional region between low-mass helium white dwarfs ($< 0.46\,M_\odot$) and typical massive CO white dwarfs ($0.7\text{–}0.9\,M_\odot$). A precise measurement of the companion mass would make this system a crucial test case for determining whether this apparent mass gap is genuine.

\begin{figure*}
	\includegraphics[width=\linewidth]{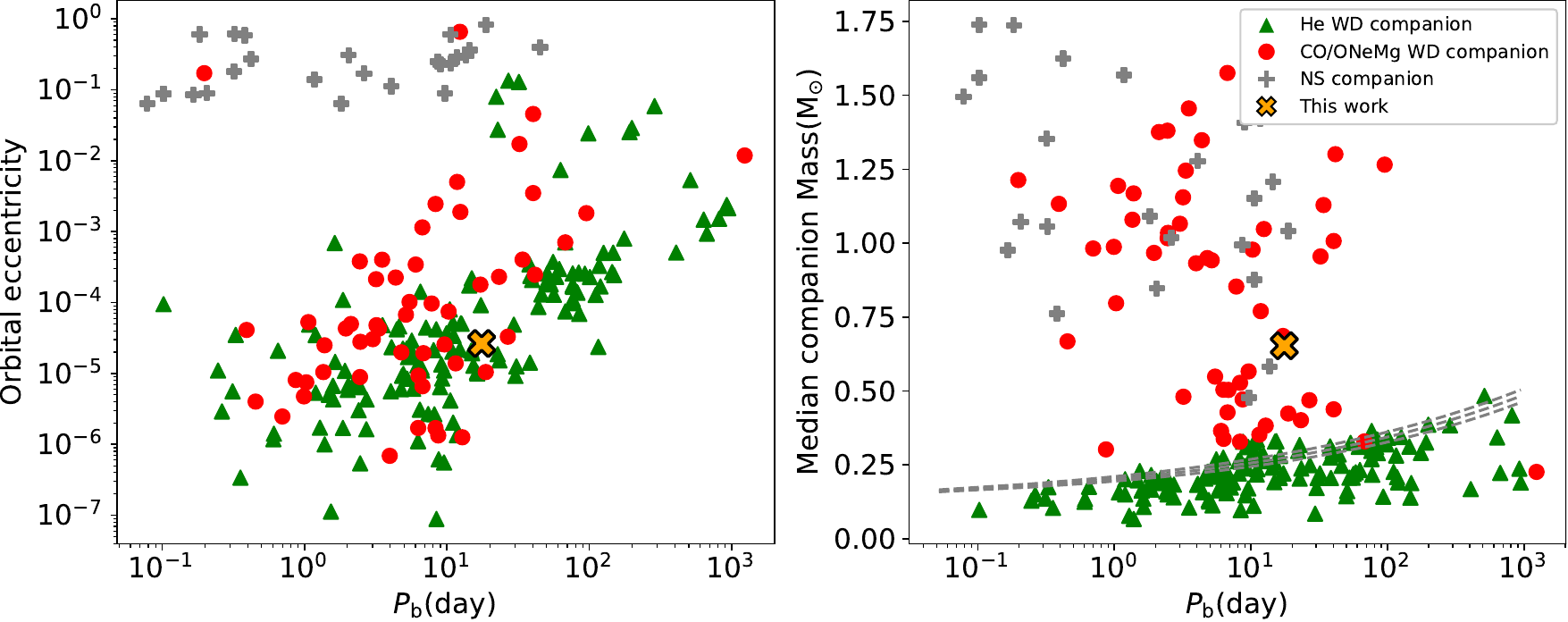}
    \caption{The left and right panels display the relationships between orbital period and, respectively, eccentricity and median companion mass for known binary systems. The data are sourced from the ATNF Pulsar Catalogue (version 2.8.1,~\citeauthor{2004IAUS..218..139H} \citeyear{2004IAUS..218..139H}; \citeauthor{2005AJ....129.1993M} \citeyear{2005AJ....129.1993M})). The dashed curve in the right panel represents the numerical results for MSP–He WD systems from~\citet{1999A&A...350..928T}.}
    \label{pb}
\end{figure*}

Figure~\ref{pb} shows orbital eccentricity (left panel) and median companion mass (right panel) as functions of orbital period for known binary systems. The left panel reveals that systems containing helium white dwarf (He WD) and carbon-oxygen/oxygen-neon-magnesium white dwarf (CO/ONeMg WD) companions exhibit broad and overlapping eccentricity distributions, with both populations showing an overall increase in orbital eccentricity as a function of orbital period in the log–log plane~\citep{1992RSPTA.341...39P,2018ApJ...864...30H}. He WD systems approximately follow the theoretical relation \(e \propto P_b\), whereas CO/ONeMg WD systems display greater scatter. For systems with relatively massive CO/ONeMg WD companions, ~\citet{2024MNRAS.534..455C} found an empirical relation of approximately \(e \propto P_b^{3/2}\). The right panel demonstrates that the He WD systems are broadly consistent with the numerical calculations of~\citet{1999A&A...350..928T}. Based on the companion mass of PSR J1912$-$1215 and its position in the diagram, its companion is more likely to be a CO/ONeMg WD.

According to the standard evolutionary model, the initially more massive star in a millisecond pulsar binary first evolves into a neutron star. It is subsequently "recycled" to millisecond spin periods by accreting matter and angular momentum from its evolving companion, a process accompanied by X-ray emission~\citep{1982Natur.300..728A,1991PhR...203....1B}. Low-mass X-ray binaries generally undergo prolonged mass transfer, producing pulsars with spin periods of a few milliseconds, whereas intermediate-mass X-ray binaries evolve more rapidly, limiting the duration of accretion and typically producing pulsars with spin periods of tens of milliseconds.

PSR J1614$-$2230 consists of a fully recycled millisecond pulsar with a spin period of only 3.15 ms and a massive CO WD companion. Evolutionary simulations by ~\cite{2011MNRAS.416.2130T,2012MNRAS.425.1601T} suggest that the system most likely originated from an intermediate-mass X-ray binary with an initial donor mass of approximately \(4.5\,M_\odot\). The relatively long-lived and stable Case A Roche-lobe overflow, especially its final phase AB lasting approximately 10 Myr, allowed the neutron star to accrete enough mass and angular momentum to be spun up to a period of 3.15 ms. Combined analyses of the binary and spin evolution rule out the Case C common-envelope channel because it cannot provide sufficient accretion to fully recycle the neutron star, whereas the Case BB channel cannot release enough orbital energy to eject the donor envelope. Therefore, Case A evolution is considered essential for the formation of this system.

PSR J1912$-$1215 may share the intermediate-mass X-ray binary origin proposed for PSR J1614$-$2230. Long-lived, stable Case A Roche-lobe overflow could have supplied enough angular momentum to produce its 5.29-ms spin. Testing this possibility requires a detailed evolutionary calculation. Longer timing may detect Shapiro delay~\citep{2023MNRAS.520.1789S}, constraining companion mass and inclination and, through the measured mass function, both component masses. Other post-Keplerian effects are less accessible in this wide, nearly circular orbit ($P_b\simeq17.5$\,d): periastron advance and Einstein delay are small, and gravitational-wave damping of $P_b$ is negligible over 3.9 yr. Shapiro delay offers the most promising mass constraint. Version 2.8.1 of the ATNF Pulsar Catalogue lists only about 20 such systems outside globular clusters.

\section{Discussion and Conclusions}
\label{sec:conclusion}
APOLLO-Timing uses a common spin and orbital model to connect observations whose relative pulse counts are unknown. Orbital delays restrict the possible rotation counts across gaps, and those counts constrain the orbit through the accumulated phase. The phase--orbit lattice retains these coupled alternatives while the continuous timing parameters are inferred. Combining the within-session phase measurements in this way allows separated observations to constrain an orbit before a coherent ephemeris has been established. Candidates are followed to zero added phase noise and checked with the full timing model at fixed pulse numbers.

APOLLO-Timing connected 96 TOAs of PSR J1912$-$1215 over 3.94 yr, grouped from 24 scans into 23 sessions, without supplied orbital parameters or a binary component. The full data yielded a unique best-supported solution with $\chi^2=188.01$, reduced $\chi^2=2.19$, and weighted rms $5.08\,\mu{\rm s}$. Shorter spans gave nonmonotonic recovery: the 15-, 17-, 18-, and 20- to 23-session runs accepted only the reference, the 16-session run accepted an alternative, and the second 19-session fit failed the checks. At 12--14 sessions, accepted paths differed from the reference despite smaller residuals. Removing sessions changes both the baseline and orbital-phase coverage, so these tests define no universal minimum observation count. Small residuals alone cannot identify the pulse count.

The 5.29-ms pulsar has $P_b=17.526$ days, a projected semimajor axis of 23.448 lt-s, and $e\simeq2.65\times10^{-5}$. For a pulsar mass of $1.4\,M_\odot$ and inclination of $60^\circ$, the inferred $0.67\,M_\odot$ companion favors a CO/ONeMg white dwarf. It lies between typical He white dwarfs ($<0.46\,M_\odot$) and massive CO white dwarfs ($0.7\text{--}0.9\,M_\odot$), making a direct mass measurement relevant to the proposed white-dwarf mass gap. Stable Case A mass transfer from an intermediate-mass X-ray binary is a possible origin. Shapiro-delay constraints on the component masses could test this interpretation.

For the solution in Table~\ref{tab:t1}, the complete 23-session calculation, from blind orbital initialization through PINT certification, required 914.0 s (15.2 min) and a maximum resident memory of 255 MiB. The calculations ran on a Ryzen 9 9950X3D host, with eight CPUs visible to the execution environment and single-threaded numerical libraries. The original chronological-prefix experiment in Table~\ref{tab:session-recovery} took 597.9--922.9 s per run. The computational cost for other systems will depend on their observing coverage and the search support.

The runtime measured for J1912$-$1215 makes repeated blind searches feasible as timing observations accumulate. In the follow-up of large radio pulsar surveys with single-dish telescopes and interferometric arrays, such searches can test whether new observations resolve competing orbital and rotation-count solutions. The resulting timing solutions yield mass functions and eccentricities that constrain companion properties and the mass-transfer histories of newly discovered binary systems.

\section*{Acknowledgments}

We thank the researchers and students in the pulsar group at Xinjiang Astronomical Observatory, and Zhichen Pan and his students at the National Astronomical Observatories, Chinese Academy of Sciences, for testing APOLLO-Timing in preparation for its open-source release.

This work is supported by the National Natural Science Foundation of China Grants (No. 12288102, 12273100), the Specialized Research Fund for State Key Laboratory of Radio Astronomy and Technology (SDQZ-KF-A2615), the Tianshan Talents Program (No. 2023TSYCTD0013), Tianshan Talents Innovation Team, the Tianchi Talents Program (Young Doctor Program) and the CAS project (No. JZHKYPT-2021-06). W. M. Yan is supported by the Tianshan Talents Program (No. 2024TSYCCX0073) and Technology Program of Xinjiang Uygur Autonomous Region (No. 2022A03013-4). J. T. Bai is supported by the National Natural Science Foundation of China Grants (No. 12603156), and the Startup Research Fund of Henan Academy of Science under Grants (No. 20251845001).
This work is also supported by the National Natural Science Foundation of China (Nos. 11988101 and T2241020) and the Tianshan Talents Program (No. 2022TSYCLJ0059).

This work made use of the data from the Five-hundred-meter Aperture Spherical radio Telescope (FAST; \url{https://cstr.cn/31116.02.FAST}). FAST is a Chinese national megascience facility, operated by the National Astronomical Observatories, Chinese Academy of Sciences.

\bibliography{sample63}{}

@ARTICLE{1989ApJ...345..434T,
       author = {{Taylor}, J.~H. and {Weisberg}, J.~M.},
        title = "{Further Experimental Tests of Relativistic Gravity Using the Binary Pulsar PSR 1913+16}",
      journal = {\apj},
         year = 1989,
        month = oct,
       volume = {345},
        pages = {434},
          doi = {10.1086/167917},
       adsurl = {https://ui.adsabs.harvard.edu/abs/1989ApJ...345..434T}
}

@ARTICLE{2014ApJ...787...82F,
       author = {{Fonseca}, Emmanuel and {Stairs}, Ingrid H. and {Thorsett}, Stephen E.},
        title = "{A Comprehensive Study of Relativistic Gravity Using PSR B1534+12}",
      journal = {\apj},
         year = 2014,
        month = may,
       volume = {787},
       number = {1},
          eid = {82},
        pages = {82},
          doi = {10.1088/0004-637X/787/1/82},
archivePrefix = {arXiv},
       eprint = {1402.4836},
 primaryClass = {astro-ph.HE},
       adsurl = {https://ui.adsabs.harvard.edu/abs/2014ApJ...787...82F}
}

@ARTICLE{2021PhRvX..11d1050K,
       author = {{Kramer}, M. and {Stairs}, I.~H. and {Manchester}, R.~N. and {Wex}, N. and {Deller}, A.~T. and {Coles}, W.~A. and {Ali}, M. and {Burgay}, M. and {Camilo}, F. and {Cognard}, I. and {Damour}, T. and {Desvignes}, G. and {Ferdman}, R.~D. and {Freire}, P.~C.~C. and {Grondin}, S. and {Guillemot}, L. and {Hobbs}, G.~B. and {Janssen}, G. and {Karuppusamy}, R. and {Lorimer}, D.~R. and {Lyne}, A.~G. and {McKee}, J.~W. and {McLaughlin}, M. and {M{\"u}nch}, L.~E. and {Perera}, B.~B.~P. and {Pol}, N. and {Possenti}, A. and {Sarkissian}, J. and {Stappers}, B.~W. and {Theureau}, G.},
        title = "{Strong-Field Gravity Tests with the Double Pulsar}",
      journal = {Physical Review X},
         year = 2021,
        month = oct,
       volume = {11},
       number = {4},
          eid = {041050},
        pages = {041050},
          doi = {10.1103/PhysRevX.11.041050},
archivePrefix = {arXiv},
       eprint = {2112.06795},
 primaryClass = {astro-ph.HE},
       adsurl = {https://ui.adsabs.harvard.edu/abs/2021PhRvX..11d1050K}
}

@ARTICLE{1974CRASM.279..971D,
       author = {{Damour}, T. and {Ruffini}, R.},
        title = "{Sur certaines v{\'e}rifications nouvelles de la Relativit{\'e} G{\'e}n{\'e}rale rendues possibles par la d{\'e}couverte d'un pulsar member d'un syst{\`e}me binaireSur certaines v{\'e}rifications nouvelles de la Relativit{\'e} G{\'e}n{\'e}rale rendues possibles par la d{\'e}couverte d'un pulsar member d'un syst{\`e}me binaireCertain new verifications of general relativity made possible by the discovery of a pulsar belonging to a binary system.}",
      journal = {Academie des Sciences Paris Comptes Rendus Serie Sciences Mathematiques},
         year = 1974,
        month = dec,
       volume = {279},
       number = {26},
        pages = {971-973},
       adsurl = {https://ui.adsabs.harvard.edu/abs/1974CRASM.279..971D}
}

@ARTICLE{1975ApJ...199L..25B,
       author = {{Barker}, B.~M. and {O'Connell}, R.~F.},
        title = "{Relativistic effects in the binary pulsar PSR 1913+16.}",
      journal = {\apjl},
         year = 1975,
        month = jul,
       volume = {199},
        pages = {L25},
          doi = {10.1086/181840},
       adsurl = {https://ui.adsabs.harvard.edu/abs/1975ApJ...199L..25B}
}

@ARTICLE{1998ApJ...509..856K,
       author = {{Kramer}, Michael},
        title = "{Determination of the Geometry of the PSR B1913+16 System by Geodetic Precession}",
      journal = {\apj},
         year = 1998,
        month = dec,
       volume = {509},
       number = {2},
        pages = {856-860},
          doi = {10.1086/306535},
archivePrefix = {arXiv},
       eprint = {astro-ph/9808127},
 primaryClass = {astro-ph},
       adsurl = {https://ui.adsabs.harvard.edu/abs/1998ApJ...509..856K}
}

@ARTICLE{2019ApJ...873L..15V,
       author = {{Venkatraman Krishnan}, V. and {Bailes}, M. and {van Straten}, W. and {Keane}, E.~F. and {Kramer}, M. and {Bhat}, N.~D.~R. and {Flynn}, C. and {Os{\l}owski}, S.},
        title = "{Relativistic Spin Precession in the Binary PSR J1141-6545}",
      journal = {\apjl},
         year = 2019,
        month = mar,
       volume = {873},
       number = {2},
          eid = {L15},
        pages = {L15},
          doi = {10.3847/2041-8213/ab0a03},
archivePrefix = {arXiv},
       eprint = {1902.09112},
 primaryClass = {astro-ph.HE},
       adsurl = {https://ui.adsabs.harvard.edu/abs/2019ApJ...873L..15V}
}

@ARTICLE{2023MNRAS.523.5064C,
       author = {{Cameron}, A.~D. and {Bailes}, M. and {Champion}, D.~J. and {Freire}, P.~C.~C. and {Kramer}, M. and {McLaughlin}, M.~A. and {Ng}, C. and {Possenti}, A. and {Ridolfi}, A. and {Tauris}, T.~M. and {Wahl}, H.~M. and {Wex}, N.},
        title = "{New constraints on the kinematic, relativistic, and evolutionary properties of the PSR J1757-1854 double neutron star system}",
      journal = {\mnras},
         year = 2023,
        month = aug,
       volume = {523},
       number = {4},
        pages = {5064-5085},
          doi = {10.1093/mnras/stad1712},
archivePrefix = {arXiv},
       eprint = {2305.14733},
 primaryClass = {astro-ph.HE},
       adsurl = {https://ui.adsabs.harvard.edu/abs/2023MNRAS.523.5064C}
}

@ARTICLE{2024ApJ...964L...7Z,
       author = {{Zhao}, D. and {Wang}, N. and {Yuan}, J.~P. and {Li}, D. and {Wang}, P. and {Xue}, M.~Y. and {Zhu}, W.~W. and {Miao}, C.~C. and {Yan}, W.~M. and {Wang}, J.~B. and {Yao}, J.~M. and {Wu}, Q.~D. and {Wang}, S.~Q. and {Sun}, S.~N. and {Kou}, F.~F. and {Chen}, Y.~T. and {Dang}, S.~J. and {Feng}, Y. and {Liu}, Z.~J. and {Miao}, X.~L. and {Meng}, L.~Q. and {Yuan}, M. and {Niu}, C.~H. and {Niu}, J.~R. and {Qian}, L. and {Wang}, S. and {Xie}, X.~Y. and {Xiao}, Y.~F. and {Yue}, Y.~L. and {You}, S.~P. and {Yu}, X.~H. and {Zhao}, R.~S. and {Yuen}, R. and {Zhou}, X. and {Zhang}, L. and {Wang}, Y.~B. and {Wu}, J.~F. and {Gan}, Z.~Y. and {Sun}, Z.~Y. and {Wang}, C.~J.},
        title = "{A Relativistic Double Neutron Star Binary PSR J1846-0513}",
      journal = {\apjl},
         year = 2024,
        month = mar,
       volume = {964},
       number = {1},
          eid = {L7},
        pages = {L7},
          doi = {10.3847/2041-8213/ad2fb3},
       adsurl = {https://ui.adsabs.harvard.edu/abs/2024ApJ...964L...7Z}
}

@article{2024ApJ...966...46M,
  author = {{Meng}, Lingqi and {Zhu}, Weiwei and {Kramer}, Michael and {Miao}, Xueli and {Desvignes}, Gregory and {Shao}, Lijing and others},
  title = {{The Relativistic Spin Precession in the Compact Double Neutron Star System PSR J1946+2052}},
  journal = {The Astrophysical Journal},
  year = {2024},
  volume = {966},
  pages = {46},
  doi = {10.3847/1538-4357/ad381c},
  eprint = {2403.17828},
  archivePrefix = {arXiv}
}

@ARTICLE{2016ARA&A..54..401O,
       author = {{{\"O}zel}, Feryal and {Freire}, Paulo},
        title = "{Masses, Radii, and the Equation of State of Neutron Stars}",
      journal = {\araa},
         year = 2016,
        month = sep,
       volume = {54},
        pages = {401-440},
          doi = {10.1146/annurev-astro-081915-023322},
archivePrefix = {arXiv},
       eprint = {1603.02698},
 primaryClass = {astro-ph.HE},
       adsurl = {https://ui.adsabs.harvard.edu/abs/2016ARA&A..54..401O}
}

@ARTICLE{2023ApJ...951L...8A,
       author = {{Agazie}, Gabriella and {Anumarlapudi}, Akash and {Archibald}, Anne M. and {Arzoumanian}, Zaven and {Baker}, Paul T. and {B{\'e}csy}, Bence and {Blecha}, Laura and {Brazier}, Adam and {Brook}, Paul R. and {Burke-Spolaor}, Sarah and {Burnette}, Rand and {Case}, Robin and {Charisi}, Maria and {Chatterjee}, Shami and {Chatziioannou}, Katerina and {Cheeseboro}, Belinda D. and {Chen}, Siyuan and {Cohen}, Tyler and {Cordes}, James M. and {Cornish}, Neil J. and {Crawford}, Fronefield and {Cromartie}, H. Thankful and {Crowter}, Kathryn and {Cutler}, Curt J. and {Decesar}, Megan E. and {Degan}, Dallas and {Demorest}, Paul B. and {Deng}, Heling and {Dolch}, Timothy and {Drachler}, Brendan and {Ellis}, Justin A. and {Ferrara}, Elizabeth C. and {Fiore}, William and {Fonseca}, Emmanuel and {Freedman}, Gabriel E. and {Garver-Daniels}, Nate and {Gentile}, Peter A. and {Gersbach}, Kyle A. and {Glaser}, Joseph and {Good}, Deborah C. and {G{\"u}ltekin}, Kayhan and {Hazboun}, Jeffrey S. and {Hourihane}, Sophie and {Islo}, Kristina and {Jennings}, Ross J. and {Johnson}, Aaron D. and {Jones}, Megan L. and {Kaiser}, Andrew R. and {Kaplan}, David L. and {Kelley}, Luke Zoltan and {Kerr}, Matthew and {Key}, Joey S. and {Klein}, Tonia C. and {Laal}, Nima and {Lam}, Michael T. and {Lamb}, William G. and {Lazio}, T. Joseph W. and {Lewandowska}, Natalia and {Littenberg}, Tyson B. and {Liu}, Tingting and {Lommen}, Andrea and {Lorimer}, Duncan R. and {Luo}, Jing and {Lynch}, Ryan S. and {Ma}, Chung-Pei and {Madison}, Dustin R. and {Mattson}, Margaret A. and {McEwen}, Alexander and {McKee}, James W. and {McLaughlin}, Maura A. and {McMann}, Natasha and {Meyers}, Bradley W. and {Meyers}, Patrick M. and {Mingarelli}, Chiara M.~F. and {Mitridate}, Andrea and {Natarajan}, Priyamvada and {Ng}, Cherry and {Nice}, David J. and {Ocker}, Stella Koch and {Olum}, Ken D. and {Pennucci}, Timothy T. and {Perera}, Benetge B.~P. and {Petrov}, Polina and {Pol}, Nihan S. and {Radovan}, Henri A. and {Ransom}, Scott M. and {Ray}, Paul S. and {Romano}, Joseph D. and {Sardesai}, Shashwat C. and {Schmiedekamp}, Ann and {Schmiedekamp}, Carl and {Schmitz}, Kai and {Schult}, Levi and {Shapiro-Albert}, Brent J. and {Siemens}, Xavier and {Simon}, Joseph and {Siwek}, Magdalena S. and {Stairs}, Ingrid H. and {Stinebring}, Daniel R. and {Stovall}, Kevin and {Sun}, Jerry P. and {Susobhanan}, Abhimanyu and {Swiggum}, Joseph K. and {Taylor}, Jacob and {Taylor}, Stephen R. and {Turner}, Jacob E. and {Unal}, Caner and {Vallisneri}, Michele and {van Haasteren}, Rutger and {Vigeland}, Sarah J. and {Wahl}, Haley M. and {Wang}, Qiaohong and {Witt}, Caitlin A. and {Young}, Olivia and {Nanograv Collaboration}},
        title = "{The NANOGrav 15 yr Data Set: Evidence for a Gravitational-wave Background}",
      journal = {\apjl},
         year = 2023,
        month = jul,
       volume = {951},
       number = {1},
          eid = {L8},
        pages = {L8},
          doi = {10.3847/2041-8213/acdac6},
archivePrefix = {arXiv},
       eprint = {2306.16213},
 primaryClass = {astro-ph.HE},
       adsurl = {https://ui.adsabs.harvard.edu/abs/2023ApJ...951L...8A}
}

@ARTICLE{2023A&A...678A..50E,
       author = {{EPTA Collaboration} and {InPTA Collaboration} and {Antoniadis}, J. and {Arumugam}, P. and {Arumugam}, S. and {Babak}, S. and {Bagchi}, M. and {Bak Nielsen}, A. -S. and {Bassa}, C.~G. and {Bathula}, A. and {Berthereau}, A. and {Bonetti}, M. and {Bortolas}, E. and {Brook}, P.~R. and {Burgay}, M. and {Caballero}, R.~N. and {Chalumeau}, A. and {Champion}, D.~J. and {Chanlaridis}, S. and {Chen}, S. and {Cognard}, I. and {Dandapat}, S. and {Deb}, D. and {Desai}, S. and {Desvignes}, G. and {Dhanda-Batra}, N. and {Dwivedi}, C. and {Falxa}, M. and {Ferdman}, R.~D. and {Franchini}, A. and {Gair}, J.~R. and {Goncharov}, B. and {Gopakumar}, A. and {Graikou}, E. and {Grie{\ss}meier}, J. -M. and {Guillemot}, L. and {Guo}, Y.~J. and {Gupta}, Y. and {Hisano}, S. and {Hu}, H. and {Iraci}, F. and {Izquierdo-Villalba}, D. and {Jang}, J. and {Jawor}, J. and {Janssen}, G.~H. and {Jessner}, A. and {Joshi}, B.~C. and {Kareem}, F. and {Karuppusamy}, R. and {Keane}, E.~F. and {Keith}, M.~J. and {Kharbanda}, D. and {Kikunaga}, T. and {Kolhe}, N. and {Kramer}, M. and {Krishnakumar}, M.~A. and {Lackeos}, K. and {Lee}, K.~J. and {Liu}, K. and {Liu}, Y. and {Lyne}, A.~G. and {McKee}, J.~W. and {Maan}, Y. and {Main}, R.~A. and {Mickaliger}, M.~B. and {Ni{\c{t}}u}, I.~C. and {Nobleson}, K. and {Paladi}, A.~K. and {Parthasarathy}, A. and {Perera}, B.~B.~P. and {Perrodin}, D. and {Petiteau}, A. and {Porayko}, N.~K. and {Possenti}, A. and {Prabu}, T. and {Quelquejay Leclere}, H. and {Rana}, P. and {Samajdar}, A. and {Sanidas}, S.~A. and {Sesana}, A. and {Shaifullah}, G. and {Singha}, J. and {Speri}, L. and {Spiewak}, R. and {Srivastava}, A. and {Stappers}, B.~W. and {Surnis}, M. and {Susarla}, S.~C. and {Susobhanan}, A. and {Takahashi}, K. and {Tarafdar}, P. and {Theureau}, G. and {Tiburzi}, C. and {van der Wateren}, E. and {Vecchio}, A. and {Venkatraman Krishnan}, V. and {Verbiest}, J.~P.~W. and {Wang}, J. and {Wang}, L. and {Wu}, Z.},
        title = "{The second data release from the European Pulsar Timing Array. III. Search for gravitational wave signals}",
      journal = {\aap},
         year = 2023,
        month = oct,
       volume = {678},
          eid = {A50},
        pages = {A50},
          doi = {10.1051/0004-6361/202346844},
archivePrefix = {arXiv},
       eprint = {2306.16214},
 primaryClass = {astro-ph.HE},
       adsurl = {https://ui.adsabs.harvard.edu/abs/2023A&A...678A..50E}
}

@ARTICLE{2023ApJ...951L...6R,
       author = {{Reardon}, Daniel J. and {Zic}, Andrew and {Shannon}, Ryan M. and {Hobbs}, George B. and {Bailes}, Matthew and {Di Marco}, Valentina and {Kapur}, Agastya and {Rogers}, Axl F. and {Thrane}, Eric and {Askew}, Jacob and {Bhat}, N.~D. Ramesh and {Cameron}, Andrew and {Cury{\l}o}, Ma{\l}gorzata and {Coles}, William A. and {Dai}, Shi and {Goncharov}, Boris and {Kerr}, Matthew and {Kulkarni}, Atharva and {Levin}, Yuri and {Lower}, Marcus E. and {Manchester}, Richard N. and {Mandow}, Rami and {Miles}, Matthew T. and {Nathan}, Rowina S. and {Os{\l}owski}, Stefan and {Russell}, Christopher J. and {Spiewak}, Ren{\'e}e and {Zhang}, Songbo and {Zhu}, Xing-Jiang},
        title = "{Search for an Isotropic Gravitational-wave Background with the Parkes Pulsar Timing Array}",
      journal = {\apjl},
         year = 2023,
        month = jul,
       volume = {951},
       number = {1},
          eid = {L6},
        pages = {L6},
          doi = {10.3847/2041-8213/acdd02},
archivePrefix = {arXiv},
       eprint = {2306.16215},
 primaryClass = {astro-ph.HE},
       adsurl = {https://ui.adsabs.harvard.edu/abs/2023ApJ...951L...6R}
}

@ARTICLE{2023RAA....23g5024X,
       author = {{Xu}, Heng and {Chen}, Siyuan and {Guo}, Yanjun and {Jiang}, Jinchen and {Wang}, Bojun and {Xu}, Jiangwei and {Xue}, Zihan and {Nicolas Caballero}, R. and {Yuan}, Jianping and {Xu}, Yonghua and {Wang}, Jingbo and {Hao}, Longfei and {Luo}, Jingtao and {Lee}, Kejia and {Han}, Jinlin and {Jiang}, Peng and {Shen}, Zhiqiang and {Wang}, Min and {Wang}, Na and {Xu}, Renxin and {Wu}, Xiangping and {Manchester}, Richard and {Qian}, Lei and {Guan}, Xin and {Huang}, Menglin and {Sun}, Chun and {Zhu}, Yan},
        title = "{Searching for the Nano-Hertz Stochastic Gravitational Wave Background with the Chinese Pulsar Timing Array Data Release I}",
      journal = {Research in Astronomy and Astrophysics},
         year = 2023,
        month = jul,
       volume = {23},
       number = {7},
          eid = {075024},
        pages = {075024},
          doi = {10.1088/1674-4527/acdfa5},
archivePrefix = {arXiv},
       eprint = {2306.16216},
 primaryClass = {astro-ph.HE},
       adsurl = {https://ui.adsabs.harvard.edu/abs/2023RAA....23g5024X}
}

@ARTICLE{2024ResPh..6107719V,
       author = {{Verbiest}, Joris P.~W. and {Vigeland}, Sarah J. and {Porayko}, Nataliya K. and {Chen}, Siyuan and {Reardon}, Daniel J.},
        title = "{Status report on global pulsar-timing-array efforts to detect gravitational waves}",
      journal = {Results in Physics},
         year = 2024,
        month = jun,
       volume = {61},
          eid = {107719},
        pages = {107719},
          doi = {10.1016/j.rinp.2024.107719},
archivePrefix = {arXiv},
       eprint = {2404.19529},
 primaryClass = {astro-ph.HE},
       adsurl = {https://ui.adsabs.harvard.edu/abs/2024ResPh..6107719V}
}

@ARTICLE{2025RAA....25a4001H,
       author = {{Han}, J.~L. and {Zhou}, D.~J. and {Wang}, C. and {Su}, W.~Q. and {Yan}, Yi and {Jing}, W.~C. and {Yang}, Z.~L. and {Wang}, P.~F. and {Wang}, T. and {Xu}, J. and {Cai}, N.~N. and {Sun}, J.~H. and {Yang}, Q.~L. and {Xu}, R.~X. and {Wang}, H.~G. and {You}, X.~P.},
        title = "{The FAST Galactic Plane Pulsar Snapshot Survey. VI. The Discovery of 473 New Pulsars}",
      journal = {Research in Astronomy and Astrophysics},
         year = 2025,
        month = jan,
       volume = {25},
       number = {1},
          eid = {014001},
        pages = {014001},
          doi = {10.1088/1674-4527/ada3b7},
archivePrefix = {arXiv},
       eprint = {2411.15961},
 primaryClass = {astro-ph.HE},
       adsurl = {https://ui.adsabs.harvard.edu/abs/2025RAA....25a4001H}
}

@ARTICLE{2026ApJ...997..210G,
       author = {{Gao}, Shi-Jie and {Shao}, Yi-Xuan and {Li}, Xiang-Dong},
        title = "{Pulsar Gleaners: Discovery of 19 Pulsars in FAST Archival Data at {\ensuremath{\mid}}b{\ensuremath{\mid}} < 5{\textdegree} and Decl. < {\ensuremath{-}}5{\textdegree}}",
      journal = {\apj},
         year = 2026,
        month = feb,
       volume = {997},
       number = {2},
          eid = {210},
        pages = {210},
          doi = {10.3847/1538-4357/ae3243},
archivePrefix = {arXiv},
       eprint = {2601.01912},
 primaryClass = {astro-ph.HE},
       adsurl = {https://ui.adsabs.harvard.edu/abs/2026ApJ...997..210G}
}

@ARTICLE{2025ApJ...982..117X,
       author = {{Xu}, Xin and {Dai}, Shi and {Zhi}, Qijun and {Bai}, Juntao and {Berteaud}, Joanna and {Calore}, Francesca and {Clavel}, Ma{\"\i}ca and {Zhu}, Weiwei and {Li}, Di and {Zhao}, Rushuang and {Xu}, Renxin and {Qiao}, Guojun},
        title = "{Discovery of 15 New Pulsars at High Galactic Latitudes with FAST}",
      journal = {\apj},
         year = 2025,
        month = apr,
       volume = {982},
       number = {2},
          eid = {117},
        pages = {117},
          doi = {10.3847/1538-4357/adbb65},
archivePrefix = {arXiv},
       eprint = {2502.20865},
 primaryClass = {astro-ph.HE},
       adsurl = {https://ui.adsabs.harvard.edu/abs/2025ApJ...982..117X}
}

@ARTICLE{2026ApJ..1001..119S,
       author = {{Sengar}, Rahul and {Anumarlapudi}, Akash and {Kaplan}, David L. and {Frail}, Dale A. and {Hyman}, Scott D. and {Polisensky}, Emil},
        title = "{Discovery of Millisecond Pulsars toward the Galactic Bulge in an Image-based Survey with MeerKAT}",
      journal = {\apj},
         year = 2026,
        month = apr,
       volume = {1001},
       number = {1},
          eid = {119},
        pages = {119},
          doi = {10.3847/1538-4357/ae4872},
archivePrefix = {arXiv},
       eprint = {2509.20614},
 primaryClass = {astro-ph.HE},
       adsurl = {https://ui.adsabs.harvard.edu/abs/2026ApJ..1001..119S}
}

@ARTICLE{2026MNRAS.545f2193P,
       author = {{Prayag}, V. and {Levin}, L. and {Geyer}, M. and {Stappers}, B.~W. and {Hurter}, H. and {Barr}, E.~D. and {Buchner}, S. and {Burgay}, M. and {Calore}, F. and {Carli}, E. and {Colom i Bernadich}, M. and {Gebauer-Werner}, L. and {Kramer}, M. and {Padmanabh}, P.~V. and {Ridolfi}, A. and {Thongmeearkom}, T. and {Turner}, J.~D. and {Venter}, C.},
        title = "{The TRAPUM Large Magellanic Cloud pulsar survey with MeerKAT ─ II. 12 new discoveries and timing solutions for 7 pulsars}",
      journal = {\mnras},
         year = 2026,
        month = feb,
       volume = {545},
       number = {4},
          eid = {staf2193},
        pages = {staf2193},
          doi = {10.1093/mnras/staf2193},
archivePrefix = {arXiv},
       eprint = {2512.12765},
 primaryClass = {astro-ph.HE},
       adsurl = {https://ui.adsabs.harvard.edu/abs/2026MNRAS.545f2193P}
}

@ARTICLE{2018MNRAS.476.4794F,
       author = {{Freire}, Paulo C.~C. and {Ridolfi}, Alessandro},
        title = "{An algorithm for determining the rotation count of pulsars}",
      journal = {\mnras},
         year = 2018,
        month = jun,
       volume = {476},
       number = {4},
        pages = {4794-4805},
          doi = {10.1093/mnras/sty524},
archivePrefix = {arXiv},
       eprint = {1802.07211},
 primaryClass = {astro-ph.IM},
       adsurl = {https://ui.adsabs.harvard.edu/abs/2018MNRAS.476.4794F}
}

@ARTICLE{2022AJ....163...84P,
       author = {{Phillips}, Camryn and {Ransom}, Scott},
        title = "{Algorithmic Pulsar Timing}",
      journal = {\aj},
         year = 2022,
        month = feb,
       volume = {163},
       number = {2},
          eid = {84},
        pages = {84},
          doi = {10.3847/1538-3881/ac403e},
       adsurl = {https://ui.adsabs.harvard.edu/abs/2022AJ....163...84P}
}

@ARTICLE{2024ApJ...964..128T,
       author = {{Taylor}, Jackson and {Ransom}, Scott and {Padmanabh}, Prajwal V.},
        title = "{Algorithmic Pulsar Timer for Binaries}",
      journal = {\apj},
         year = 2024,
        month = apr,
       volume = {964},
       number = {2},
          eid = {128},
        pages = {128},
          doi = {10.3847/1538-4357/ad1ce9},
archivePrefix = {arXiv},
       eprint = {2310.10800},
 primaryClass = {astro-ph.HE},
       adsurl = {https://ui.adsabs.harvard.edu/abs/2024ApJ...964..128T}
}

@ARTICLE{1982Natur.300..728A,
       author = {{Alpar}, M.~A. and {Cheng}, A.~F. and {Ruderman}, M.~A. and {Shaham}, J.},
        title = "{A new class of radio pulsars}",
      journal = {\nat},
         year = 1982,
        month = dec,
       volume = {300},
       number = {5894},
        pages = {728-730},
          doi = {10.1038/300728a0},
       adsurl = {https://ui.adsabs.harvard.edu/abs/1982Natur.300..728A}
}

@ARTICLE{1991PhR...203....1B,
       author = {{Bhattacharya}, D. and {van den Heuvel}, E.~P.~J.},
        title = "{Formation and evolution of binary and millisecond radio pulsars}",
      journal = {\physrep},
         year = 1991,
        month = jan,
       volume = {203},
       number = {1-2},
        pages = {1-124},
          doi = {10.1016/0370-1573(91)90064-S},
       adsurl = {https://ui.adsabs.harvard.edu/abs/1991PhR...203....1B}
}

@ARTICLE{2011MNRAS.416.2130T,
       author = {{Tauris}, T.~M. and {Langer}, N. and {Kramer}, M.},
        title = "{Formation of millisecond pulsars with CO white dwarf companions - I. PSR J1614-2230: evidence for a neutron star born massive}",
      journal = {\mnras},
         year = 2011,
        month = sep,
       volume = {416},
       number = {3},
        pages = {2130-2142},
          doi = {10.1111/j.1365-2966.2011.19189.x},
archivePrefix = {arXiv},
       eprint = {1103.4996},
 primaryClass = {astro-ph.SR},
       adsurl = {https://ui.adsabs.harvard.edu/abs/2011MNRAS.416.2130T}
}

@ARTICLE{2012MNRAS.425.1601T,
       author = {{Tauris}, T.~M. and {Langer}, N. and {Kramer}, M.},
        title = "{Formation of millisecond pulsars with CO white dwarf companions - II. Accretion, spin-up, true ages and comparison to MSPs with He white dwarf companions}",
      journal = {\mnras},
         year = 2012,
        month = sep,
       volume = {425},
       number = {3},
        pages = {1601-1627},
          doi = {10.1111/j.1365-2966.2012.21446.x},
archivePrefix = {arXiv},
       eprint = {1206.1862},
 primaryClass = {astro-ph.SR},
       adsurl = {https://ui.adsabs.harvard.edu/abs/2012MNRAS.425.1601T}
}

@ARTICLE{2023MNRAS.520.1789S,
       author = {{Shamohammadi}, M. and {Bailes}, M. and {Freire}, P.~C.~C. and {Parthasarathy}, A. and {Reardon}, D.~J. and {Shannon}, R.~M. and {Venkatraman Krishnan}, V. and {Bernadich}, M.~C. i. and {Cameron}, A.~D. and {Champion}, D.~J. and {Corongiu}, A. and {Flynn}, C. and {Geyer}, M. and {Kramer}, M. and {Miles}, M.~T. and {Possenti}, A. and {Spiewak}, R.},
        title = "{Searches for Shapiro delay in seven binary pulsars using the MeerKAT telescope}",
      journal = {\mnras},
         year = 2023,
        month = apr,
       volume = {520},
       number = {2},
        pages = {1789-1806},
          doi = {10.1093/mnras/stac3719},
archivePrefix = {arXiv},
       eprint = {2212.04051},
 primaryClass = {astro-ph.HE},
       adsurl = {https://ui.adsabs.harvard.edu/abs/2023MNRAS.520.1789S}
}

@ARTICLE{1999A&A...350..928T,
       author = {{Tauris}, Thomas M. and {Savonije}, Gerrit J.},
        title = "{Formation of millisecond pulsars. I. Evolution of low-mass X-ray binaries with P\_orb> 2 days}",
      journal = {\aap},
         year = 1999,
        month = oct,
       volume = {350},
        pages = {928-944},
          doi = {10.48550/arXiv.astro-ph/9909147},
archivePrefix = {arXiv},
       eprint = {astro-ph/9909147},
 primaryClass = {astro-ph},
       adsurl = {https://ui.adsabs.harvard.edu/abs/1999A&A...350..928T}
}

@ARTICLE{1992RSPTA.341...39P,
       author = {{Phinney}, E.~S.},
        title = "{Pulsars as Probes of Newtonian Dynamical Systems}",
      journal = {Philosophical Transactions of the Royal Society of London Series A},
         year = 1992,
        month = oct,
       volume = {341},
       number = {1660},
        pages = {39-75},
          doi = {10.1098/rsta.1992.0084},
       adsurl = {https://ui.adsabs.harvard.edu/abs/1992RSPTA.341...39P}
}

@ARTICLE{2018ApJ...864...30H,
       author = {{Hui}, C.~Y. and {Wu}, Kinwah and {Han}, Qin and {Kong}, A.~K.~H. and {Tam}, P.~H.~T.},
        title = "{On the Orbital Properties of Millisecond Pulsar Binaries}",
      journal = {\apj},
         year = 2018,
        month = sep,
       volume = {864},
       number = {1},
          eid = {30},
        pages = {30},
          doi = {10.3847/1538-4357/aad5ec},
archivePrefix = {arXiv},
       eprint = {1807.09001},
 primaryClass = {astro-ph.HE},
       adsurl = {https://ui.adsabs.harvard.edu/abs/2018ApJ...864...30H}
}

@ARTICLE{2024MNRAS.534..455C,
       author = {{Cohen}, Yair and {Ginzburg}, Sivan and {Levy}, Maya and {Bar Shalom}, Tal and {Siman Tov}, Yoav},
        title = "{White dwarf eccentricity fluctuation and dissipation by AGB convection}",
      journal = {\mnras},
         year = 2024,
        month = oct,
       volume = {534},
       number = {1},
        pages = {455-464},
          doi = {10.1093/mnras/stae2136},
archivePrefix = {arXiv},
       eprint = {2405.03745},
 primaryClass = {astro-ph.SR},
       adsurl = {https://ui.adsabs.harvard.edu/abs/2024MNRAS.534..455C}
}

@ARTICLE{2021MNRAS.508..300C,
       author = {{Cruces}, M. and {Champion}, D.~J. and {Li}, D. and {Kramer}, M. and {Zhu}, W.~W. and {Wang}, P. and {Cameron}, A.~D. and {Chen}, Y.~T. and {Hobbs}, G. and {Freire}, P.~C.~C. and {Graikou}, E. and {Krco}, M. and {Liu}, Z.~J. and {Miao}, C.~C. and {Niu}, J. and {Pan}, Z.~C. and {Qian}, L. and {Xue}, M.~Y. and {Xie}, X.~Y. and {You}, S.~P. and {Yu}, X.~H. and {Yuan}, M. and {Yue}, Y.~L. and {Zhu}, Y. and {Zhu}, Y. and {Lackeos}, K. and {Porayko}, N. and {Wongphecauxon}, J. and {Main}, R. and {Crafts Collaboration}},
        title = "{FAST early pulsar discoveries: Effelsberg follow-up}",
      journal = {\mnras},
         year = 2021,
        month = nov,
       volume = {508},
       number = {1},
        pages = {300-314},
          doi = {10.1093/mnras/stab2540},
archivePrefix = {arXiv},
       eprint = {2108.09121},
 primaryClass = {astro-ph.HE},
       adsurl = {https://ui.adsabs.harvard.edu/abs/2021MNRAS.508..300C}
}

@ARTICLE{2023MNRAS.518.1672M,
       author = {{Miao}, C.~C. and {Zhu}, W.~W. and {Li}, D. and {Freire}, P.~C.~C. and {Niu}, J.~R. and {Wang}, P. and {Yuan}, J.~P. and {Xue}, M.~Y. and {Cameron}, A.~D. and {Champion}, D.~J. and {Cruces}, M. and {Chen}, Y.~T. and {Chi}, M.~M. and {Cheng}, X.~F. and {Dang}, S.~J. and {Ding}, M.~F. and {Feng}, Y. and {Gan}, Z.~Y. and {Hobbs}, G. and {Kramer}, M. and {Liu}, Z.~J. and {Li}, Y.~X. and {Luo}, Z.~K. and {Miao}, X.~L. and {Meng}, L.~Q. and {Niu}, C.~H. and {Pan}, Z.~C. and {Qian}, L. and {Sun}, Z.~Y. and {Wang}, N. and {Wang}, S.~Q. and {Wang}, J.~B. and {Wu}, Q.~D. and {Wang}, Y.~B. and {Wang}, C.~J. and {Wang}, H.~F. and {Wang}, S. and {Xie}, X.~Y. and {Xie}, M. and {Xiao}, Y.~F. and {Yuan}, M. and {Yue}, Y.~L. and {Yao}, J.~M. and {Yan}, W.~M. and {You}, S.~P. and {Yu}, X.~H. and {Zhao}, D. and {Zhao}, R.~S. and {Zhang}, L.},
        title = "{Arecibo and FAST timing follow-up of 12 millisecond pulsars discovered in Commensal Radio Astronomy FAST Survey}",
      journal = {\mnras},
         year = 2023,
        month = jan,
       volume = {518},
       number = {2},
        pages = {1672-1682},
          doi = {10.1093/mnras/stac1305},
archivePrefix = {arXiv},
       eprint = {2205.04232},
 primaryClass = {astro-ph.HE},
       adsurl = {https://ui.adsabs.harvard.edu/abs/2023MNRAS.518.1672M}
}

@ARTICLE{2024ApJ...975...88Z,
       author = {{Zhao}, D. and {Yuan}, J.~P. and {Wang}, N. and {Li}, D. and {Wang}, P. and {Xue}, M.~Y. and {Zhu}, W.~W. and {Miao}, C.~C. and {Yan}, W.~M. and {Wang}, J.~B. and {Yao}, J.~M. and {Wu}, Q.~D. and {Wang}, S.~Q. and {Sun}, S.~N. and {Kou}, F.~F. and {Chen}, Y.~T. and {Dang}, S.~J. and {Feng}, Y. and {Liu}, Z.~J. and {Miao}, X.~L. and {Meng}, L.~Q. and {Yuan}, M. and {Niu}, C.~H. and {Niu}, J.~R. and {Qian}, L. and {Wang}, S. and {Xie}, X.~Y. and {Xiao}, Y.~F. and {Yue}, Y.~L. and {You}, S.~P. and {Yu}, X.~H. and {Zhao}, R.~S. and {Yuen}, R. and {Zhou}, X. and {Zhang}, L. and {Xie}, M. and {Li}, Y.~X. and {Wang}, Y.~B. and {Luo}, Z.~K. and {Gan}, Z.~Y. and {Sun}, Z.~Y. and {Chi}, M.~M. and {Wang}, C.~J.},
        title = "{Follow-up Timing of 12 Pulsars Discovered in Commensal Radio Astronomy FAST Survey}",
      journal = {\apj},
         year = 2024,
        month = nov,
       volume = {975},
       number = {1},
          eid = {88},
        pages = {88},
          doi = {10.3847/1538-4357/ad79f4},
archivePrefix = {arXiv},
       eprint = {2410.09518},
 primaryClass = {astro-ph.HE},
       adsurl = {https://ui.adsabs.harvard.edu/abs/2024ApJ...975...88Z}
}

@ARTICLE{2021ApJ...911...45L,
       author = {{Luo}, Jing and {Ransom}, Scott and {Demorest}, Paul and {Ray}, Paul S. and {Archibald}, Anne and {Kerr}, Matthew and {Jennings}, Ross J. and {Bachetti}, Matteo and {van Haasteren}, Rutger and {Champagne}, Chloe A. and {Colen}, Jonathan and {Phillips}, Camryn and {Zimmerman}, Josef and {Stovall}, Kevin and {Lam}, Michael T. and {Jenet}, Fredrick A.},
        title = "{PINT: A Modern Software Package for Pulsar Timing}",
      journal = {\apj},
         year = 2021,
        month = apr,
       volume = {911},
       number = {1},
          eid = {45},
        pages = {45},
          doi = {10.3847/1538-4357/abe62f},
archivePrefix = {arXiv},
       eprint = {2012.00074},
 primaryClass = {astro-ph.IM},
       adsurl = {https://ui.adsabs.harvard.edu/abs/2021ApJ...911...45L}
}

@ARTICLE{2017ApJ...835...29Y,
       author = {{Yao}, J.~M. and {Manchester}, R.~N. and {Wang}, N.},
        title = "{A New Electron-density Model for Estimation of Pulsar and FRB Distances}",
      journal = {\apj},
         year = 2017,
        month = jan,
       volume = {835},
       number = {1},
          eid = {29},
        pages = {29},
          doi = {10.3847/1538-4357/835/1/29},
archivePrefix = {arXiv},
       eprint = {1610.09448},
 primaryClass = {astro-ph.GA},
       adsurl = {https://ui.adsabs.harvard.edu/abs/2017ApJ...835...29Y}
}

@ARTICLE{2026ApJ..1002....3O,
       author = {{Ocker}, Stella Koch and {Cordes}, James M.},
        title = "{NE2025: An Updated Electron Density Model for the Galactic Interstellar Medium}",
      journal = {\apj},
         year = 2026,
        month = may,
       volume = {1002},
       number = {1},
          eid = {3},
        pages = {3},
          doi = {10.3847/1538-4357/ae5825},
archivePrefix = {arXiv},
       eprint = {2602.11838},
 primaryClass = {astro-ph.GA},
       adsurl = {https://ui.adsabs.harvard.edu/abs/2026ApJ..1002....3O}
}

@INPROCEEDINGS{2004IAUS..218..139H,
       author = {{Hobbs}, G. and {Manchester}, R. and {Teoh}, A. and {Hobbs}, M.},
        title = "{The ATNF Pulsar Catalog}",
    booktitle = {Young Neutron Stars and Their Environments},
         year = 2004,
       editor = {{Camilo}, Fernando and {Gaensler}, Bryan M.},
       series = {IAU Symposium},
       volume = {218},
        month = jan,
        pages = {139},
          doi = {10.48550/arXiv.astro-ph/0309219},
archivePrefix = {arXiv},
       eprint = {astro-ph/0309219},
 primaryClass = {astro-ph},
       adsurl = {https://ui.adsabs.harvard.edu/abs/2004IAUS..218..139H}
}

@ARTICLE{2005AJ....129.1993M,
       author = {{Manchester}, R.~N. and {Hobbs}, G.~B. and {Teoh}, A. and {Hobbs}, M.},
        title = "{The Australia Telescope National Facility Pulsar Catalogue}",
      journal = {\aj},
         year = 2005,
        month = apr,
       volume = {129},
       number = {4},
        pages = {1993-2006},
          doi = {10.1086/428488},
archivePrefix = {arXiv},
       eprint = {astro-ph/0412641},
 primaryClass = {astro-ph},
       adsurl = {https://ui.adsabs.harvard.edu/abs/2005AJ....129.1993M}
}

@ARTICLE{2004PASA...21..302H,
       author = {{Hotan}, A.~W. and {van Straten}, W. and {Manchester}, R.~N.},
        title = "{PSRCHIVE and PSRFITS: An Open Approach to Radio Pulsar Data Storage and Analysis}",
      journal = {\pasa},
         year = 2004,
        month = jan,
       volume = {21},
       number = {3},
        pages = {302-309},
          doi = {10.1071/AS04022},
archivePrefix = {arXiv},
       eprint = {astro-ph/0404549},
 primaryClass = {astro-ph},
       adsurl = {https://ui.adsabs.harvard.edu/abs/2004PASA...21..302H}
}
\bibliographystyle{aasjournal}

\end{document}